\documentclass[12pt]{article}
\usepackage{amsmath,amssymb}
\usepackage[margin=.8 in]{geometry}
\usepackage{hyperref}
\usepackage{mathtools}
\usepackage{color}
\usepackage{graphicx,float}
\usepackage{authblk}
\newcommand{\precc}{\mathrel{\prec\!\!\ast}}
\newcommand{\dd}{\text{d}}
\newcommand{\beq}{\begin{equation}}\newcommand{\eeq}{\end{equation}}
\newcommand{\bea}{\begin{eqnarray}}\newcommand{\eea}{\end{eqnarray}}

\begin{document}
\title{\bf Discrete Spacetime: a Web of Chains}
\author[1]{M. Aghili\thanks{maghili@go.olemiss.edu}}
\author[1]{L. Bombelli\thanks{bombelli@olemiss.edu}}
\author[1]{B.B. Pilgrim\thanks{bbpilgri@go.olemiss.edu}}
\affil[1]{Department of Physics and Astronomy, The University of Mississippi,
University, MS 38677-1848}
\date{January 31, 2019}
\maketitle
\vspace{10pt}
\abstract{\noindent This paper studies the distribution of chain and maximal chain lengths in a causal set. We first provide a new derivation for these distributions for a causal set uniformly embedded in Minkowski space, for various dimensionalities, which includes a correction to previously available expressions. Results of numerical simulations show a better agreement with the improved theoretical distributions. As examples of applications of those distributions, we then expand on previous work of ours regarding their use in establishing whether a causal set is manifoldlike and, if it is, finding its dimensionality. We then compare this measure of the dimension with other methods in the literature.}
\section{Introduction}
Causal set theory is an approach to quantum gravity in which the continuum model of spacetime as a Lorentzian manifold is replaced by a locally finite partially ordered set, a set $\mathcal{C}$ with a relation on $\mathcal{C}$ which is:
\begin{enumerate}
\item Irreflexive:\footnote{Partial orders are often defined as reflexive relations, such that $p\prec p$ for all $p\in\mathcal{C}$, satisfying the additional antisymmetry condition that $p\prec q \Rightarrow q\not\prec p$ for all $p,q \in\mathcal{C}$. Each reflexive order corresponds to a unique irreflexive one (and vice versa.)} $\forall p\in\mathcal{C}, \quad p\not\prec p$
\item Transitive: $\forall p,q,r\in\mathcal{C}, \quad
p\prec q\prec r \implies p\prec r$
\item Locally finite: $\forall p,q \in\mathcal{C}, \quad \#(I(p,q))<\infty$\;.
\end{enumerate}
Here, the set $I(p,q) \coloneqq \{r \mid p\prec r\prec q\}$ is the {\em interval\/} defined by the elements $p$ and $q$, and $\#(X)$ denotes the cardinality of a set $X$. We interpret the relation $p \prec q$ as a causal structure on ${\mathcal{C}}$, the discrete analog of the chronology relation $x \in I^-(y)$ on a Lorentzian manifold (this relation is a partial order if the spacetime has a well-behaved causal structure), and the interval is the discrete analog of an Alexandrov set $A(x,y):= I^+(x) \cap I^-(y)$ in the continuum. We will now define a few other terms to be used throughout this paper. A \textit{chain} is a subset of $\mathcal{C}$ all of whose elements are pairwise related; in other words, a $k$-chain is a totally ordered set of the form $\{p_1, p_2, \hdots, p_{k+1} \mid p_1\prec p_2\prec \hdots \prec p_{k+1}\}$. Two elements of ${\mathcal{C}}$ are {\em linked\/} if they are related with no elements between them; we will denote this by the relation $p\precc q :\Leftrightarrow (p\prec q)\ \&\ (I(p,q) = \emptyset)$. A \textit{path} is a maximal chain, i.e., one which cannot be lengthened by including a new element of $\mathcal C$ between any two of its elements; a $k$-path is then a set of the form $\coloneqq {\{p_1,p_2,\hdots, p_{k+1}\mid p_1\precc p_2\precc \hdots \precc p_{k+1}\}}$. 

A (large) causal set $\mathcal C$ and a Lorentzian manifold $({\mathcal M},g)$ are seen as good approximations of each other if the causal set can be faithfully embedded in $\mathcal M$ \cite{sorbom}, which for the purposes of this paper can be taken to mean that there is a embedding $f: {\mathcal C}\to {\mathcal M}$ under which the elements of $\mathcal C$ are distributed with uniform density in $\mathcal M$ and the partial orders agree in the sense that $p\prec q$ iff $f(p) \in I^-(f(q))$. The uniform density condition is necessary, at least in flat spacetime, to ensure that the embedding process does not introduce a preferred frame or other additional structure in ${\mathcal M}$.

In a previous paper \cite{paths} we showed how to calculate the path length distribution (mean number of $k$-paths as a function of $k$) for $N$-element causal sets obtained from uniformly random distributions of points inside an Alexandrov set $A(x,y)$ in 2-dimensional Minkowski space. Using that distribution and the results of numerical simulations we argued that the mean path length and width of the distribution provide a good criterion of embeddability of a causal set in flat spacetime. This paper extends those results in various ways. In Section 2 we calculate the chain length distribution for causal sets uniformly embedded in $d$-dimensional Minkowski space. Importantly, we point out a correction to previous results on the chain-length distribution in the literature, and to the method we had previously used to calculate the path-length distribution \cite{paths}, which we revisit, correct and extend to $d$ dimensions in Section 3. In Section 4 we return to the question of manifoldlikeness of causal sets, for which in the case of chains we can also calculate analytically the width of the length distribution. These results show the usefulness of these length distributions for the analysis of causal sets, and in Section 5 we compare the results obtained using these and other methods when estimating the dimensionality of a flat spacetime in which a causal set can be embedded.
\bigskip

\section{Chain Length Distributions in Minkowski Space}\label{chainlengths}

In this section we derive the chain length distribution for a causal set uniformly embedded in Minkowski spacetime. The Lorentzian manifolds associated with manifoldlike causal sets may be curved, but in sufficiently small local regions they are approximately flat, so Minkowski space is of interest both as a simple spacetime in its own right, and as a starting point for the study of the distribution of chain lengths in more general causal sets.

As a result of a uniformly random process of choosing (``sprinkling") $N$ points in a manifold $\mathcal M$ of volume $V_0$, the probability of finding $n$ of those points in a region $R\subset{\mathcal M}$ is given by the binomial distribution,
$$
P(\ell) = {N\choose \ell}\, (V(R)/V_0)^\ell\,(1-V(R)/V_0)^{N-\ell}\;,
$$
which for fixed density $\rho:= N/V_0$ in the $\ell \ll N$ limit becomes the Poisson distribution,
$$
P(\ell) \approx \text{e}^{-\rho V}(\rho V)^\ell/\ell!\;.
$$
For example, in an infinitesimal $d$-dimensional volume $\dd V = \sqrt{-g(x)}\,\dd^dx$ the probability of finding a single point is $\rho\,\dd V$,\footnote{The factor $\sqrt{-g(x)}$, in which $g(x)$ is the determinant of the metric, is needed even in Minkowski space if non-Cartesian coordinates are used.} while the probability of finding a larger number of points is a higher-order differential.

\begin{figure}[h]
\begin{center}
\includegraphics[width=.45\textwidth]{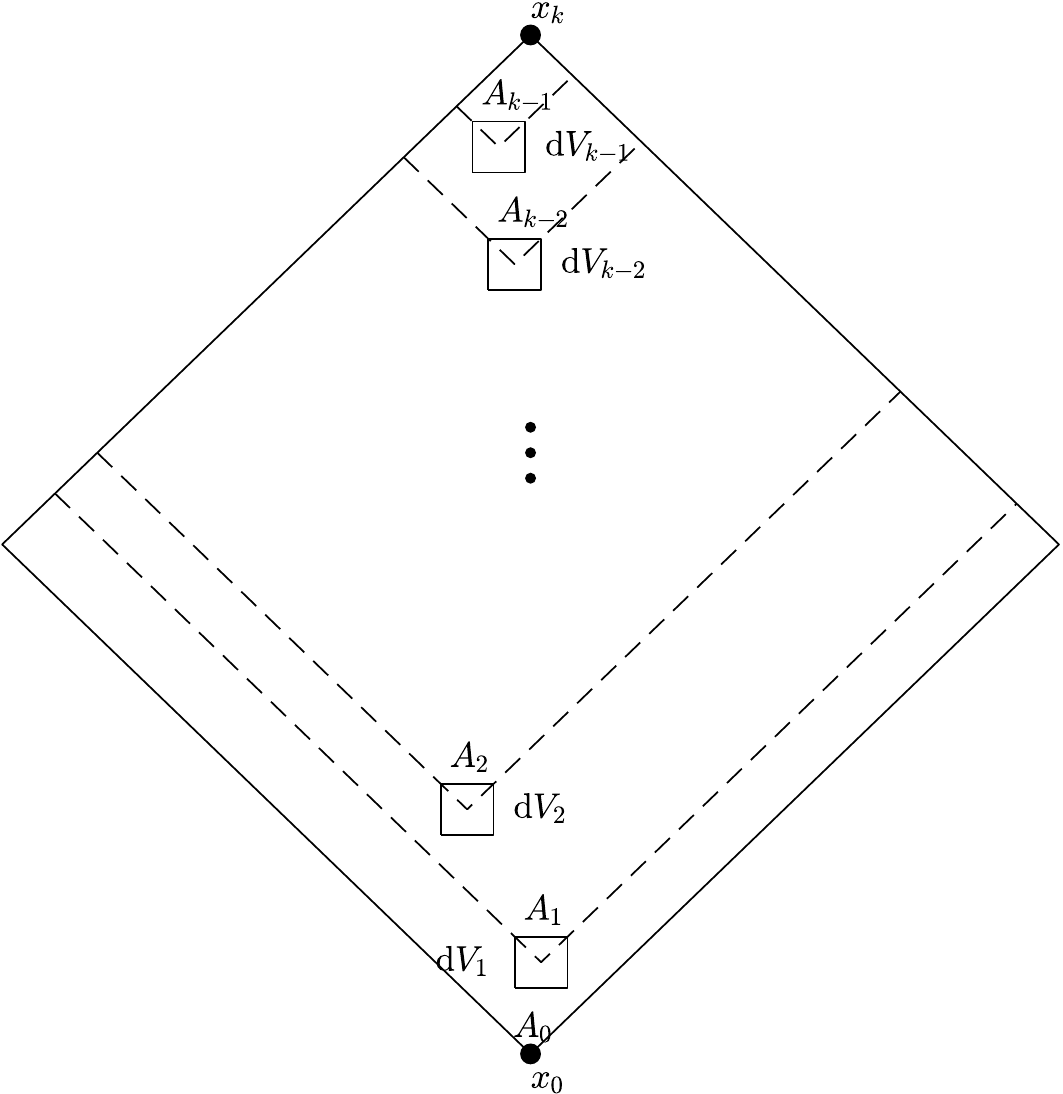}
\end{center}
\caption{A visual representation of the volumes involved in the calculation of the expected number of chains $\left<c_k\right>$ of length $k$ in a causal set obtained from points uniformly distributed inside an Alexandrov set in Minkowski spacetime.}
\label{fig:volumes}
\end{figure}

Our goal is to use as the manifold in which we sprinkle $N$ points uniformly at random an Alexandrov set ${\cal M} = A_0 = A(x_0,x_{N+1})$ in Minkowski space of volume $V_0 = V(A_0)$, and find the mean number $c_k$ of chains of length $k$ in the resulting causal set. For each particular set of locations $x_i \in A(x_0,x_{N+1})$, $i = 1, ..., k-1$, the probability of finding a $k$-chain from $x_0$ to $x_{N+1}$ through the volumes $\dd V_1$, $\dd V_2$, ..., $\dd V_{k-1}$ around $x_1$, $x_2$, ..., $x_{k-1}$ can be calculated by induction as follows. For $k = 1$ the probability is 1, and for $k = 2$ and 3 it is
\bea
   & &\dd P_{(2)}(x_1) = \rho_0\,\dd V_1
   = \frac{N}{V_0}\,\sqrt{-g(x_1)}\,\dd^dx\;, \label{P2}\\
   & &\dd P_{(3)}(x_1,x_2) = \dd P_{(2)}(x_1)\,\rho_1\,\dd V_2
   = \frac{N(N-1)}{V_0^2}\,\dd V_1\,\dd V_2\;, \label{P3}
\eea
provided that $x_2 \in I^+(x_1)$, where $\rho_0 = \rho = N/V_0$ and $\rho_1 = (N-1)/V_0$ is the density with which the remaining points are distributed given that one point is located at $x_1$ and there are only $N-1$ points left to locate; notice that these densities do not have the same value, an important observation whose consequences we will come back to below. Similarly, for each other $k$ the probability is obtained from the recursion relation
\beq
   \dd P_{(k)}^{~}(x_1,...,x_{k-1})
   = \dd P_{(k-1)}^{~}(x_1,...,x_{k-2})\,\rho_{k-2}^{~}\,\dd V_{k-1}^{~}\;, \label{Pk}
\eeq
where $\rho_i = (N-i)/V_0$ for all $i$. The expected number of $k$-chains from $x_0$ to $x_{N+1}$ through the $\dd V_i$ is then the sum, over the possible values of the number $c_k$ of $k$-chains, of $c_k$ times the probability that there are $c_k$ chains through those points. The only contributing term to that sum is the one with $c_k = 1$, so
\beq
   \left<\dd c_k(x_1,...,x_{k-1})\right>
   = \dd P_{(k)}(x_1,...,x_{k-1})\;,
\eeq
and we can therefore find the mean total number $\left<c_k\right>$ of $k$-chains from $x_0$ to $x_{N+1}$ by integrating this expression over all the $x_i$, with each point in the future of the previous one, $x_i\in I^+(x_{i-1})$,
\bea
   & &\left<c_k\right>
   = \int_{A_0} \rho_0\,\dd V_1 \int_{A_1} \rho_1\,\dd V_2 \cdots
   \int_{A_{k-2}} \rho_{k-2}\,\dd V_{k-1}\;, \notag\\
   & &= \frac{N(N-1)(N-2){\cdots}(N-(k-2))}{V^{k-1}}
   \int_{A_0}\dd V_1\int_{A_1}\dd V_2\cdots\int_{A_{k-2}}\dd V_{k-1}\;,
   \label{Eq:gchain}
\eea
where $A_i$ is short for $A(x_i,x_k)$. A guide to the eye is provided in Fig.\ \ref{fig:volumes}, where for a $k$-chain we are identifying $x_k = x_{N+1}$. After integrating and rewriting the coefficient, in Minkowski space this expression simplifies to
\beq
   \left<c_k\right>
   = \frac{N!}{(N-(k-1))!} \left(\frac{\Gamma(d+1)}{2}\right)^{\!k-2}
   \frac{\Gamma(d/2+1)\,\Gamma(d)}{\Gamma((k-1)d/2+1)\,\Gamma(kd/2)}\;,\label{chainsO}
\eeq
for all $k$.

\begin{figure}[htp]
\begin{center}
\includegraphics[width=.45\textwidth]{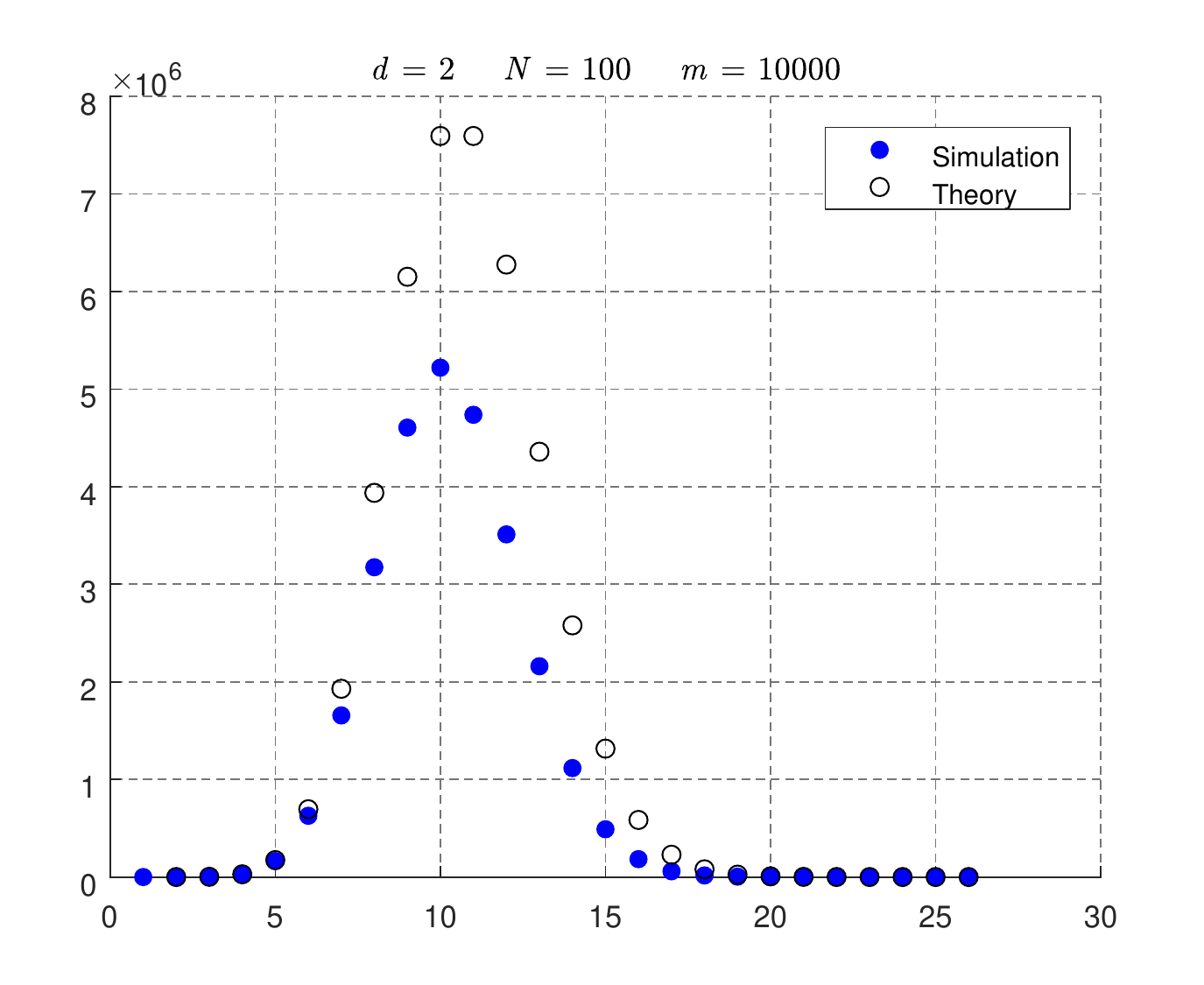}
\includegraphics[width=.45\textwidth]{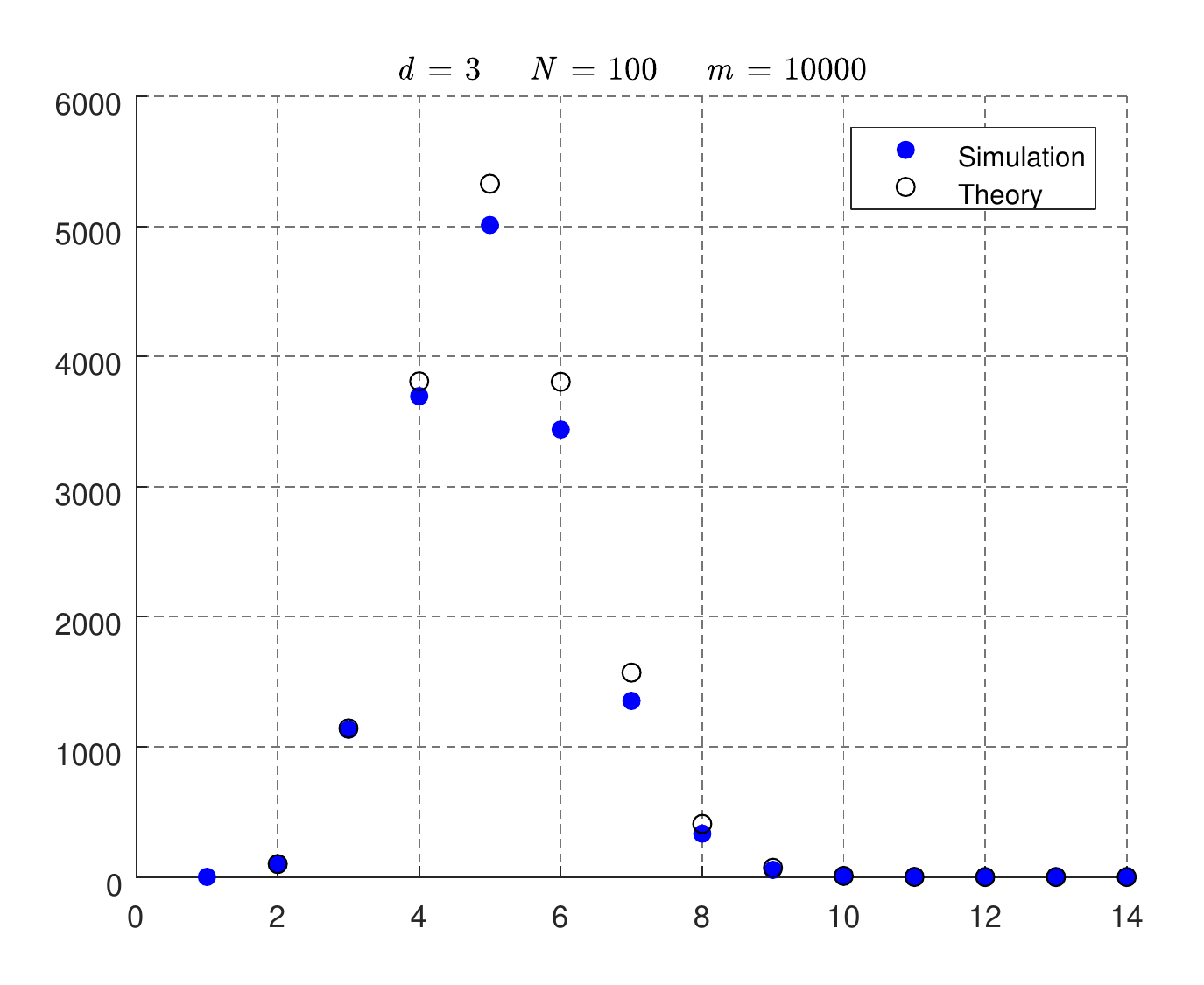}
\includegraphics[width=.45\textwidth]{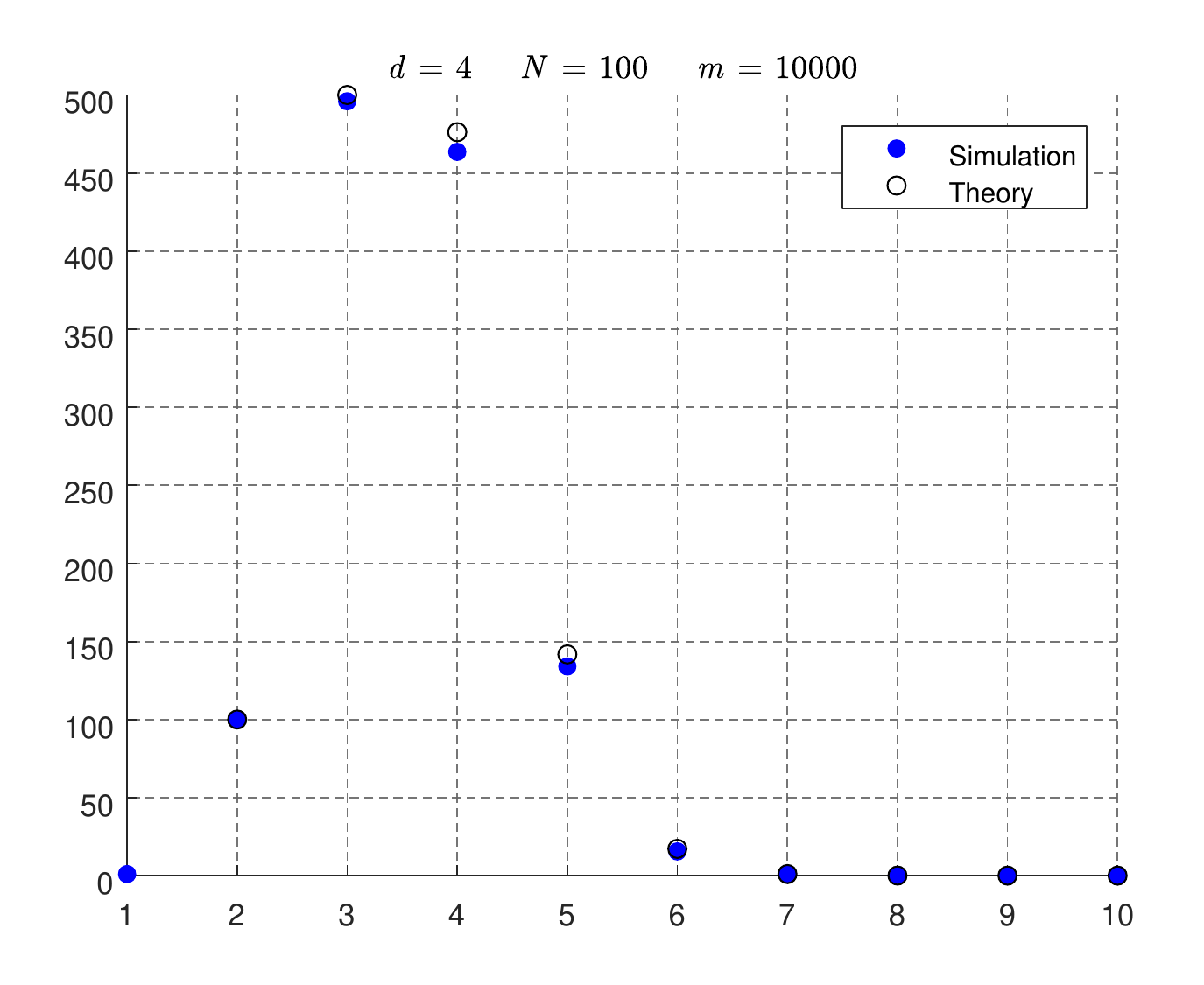}
\includegraphics[width=.45\textwidth]{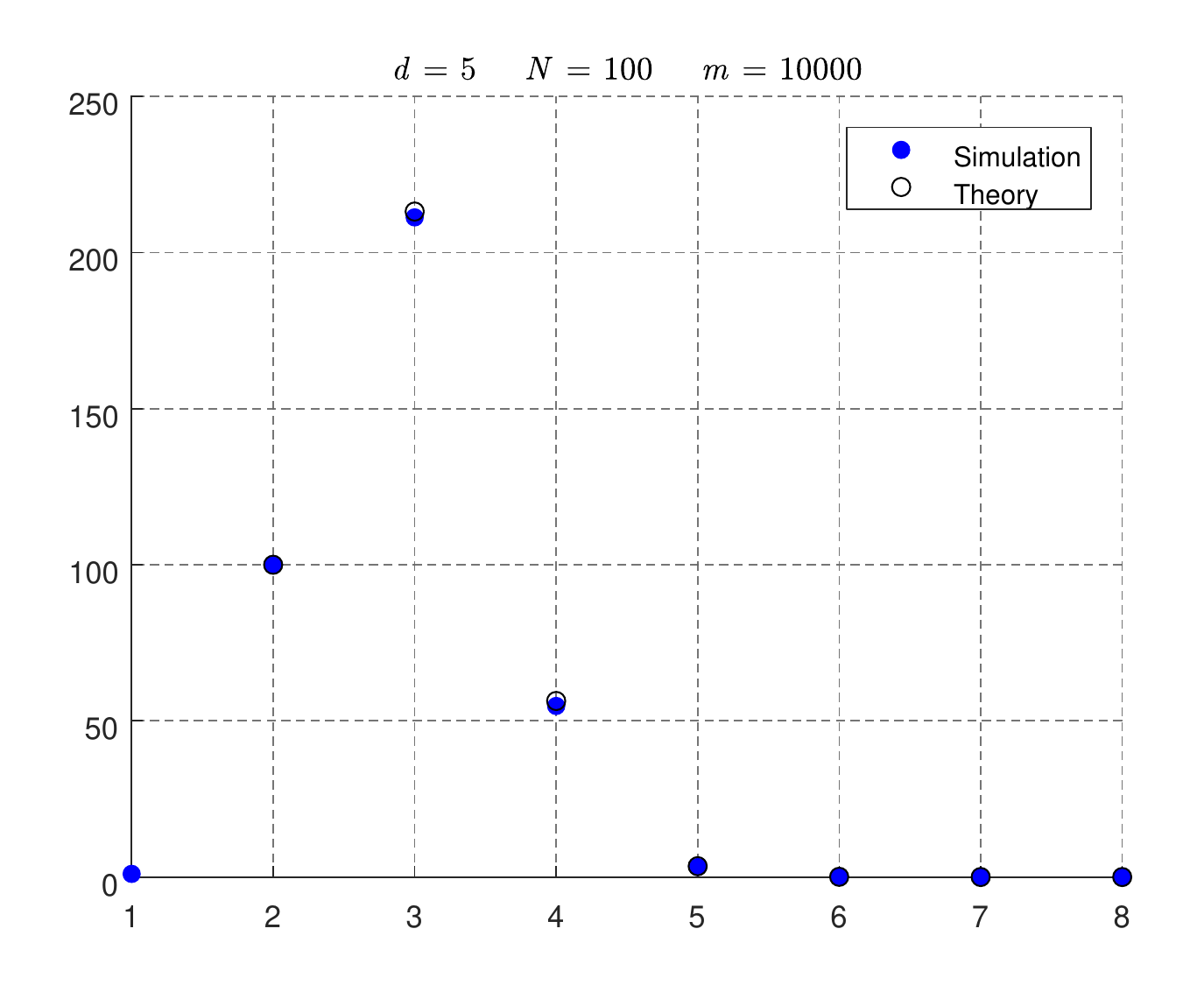}
\end{center}
\caption{Plots of the average number of $k$-chains as a function of $k$ for $m = 10,000$ simulated sprinklings of $N = 100$ points in Minkowski space of various dimensionalities, and the mean number of $k$-chains computed using the $\rho_i = \rho_0$ distribution (\ref{chainsM}).}
\label{fig:distsM}
\end{figure}

\begin{figure}[htp]
\begin{center}
\includegraphics[width=.45\textwidth]{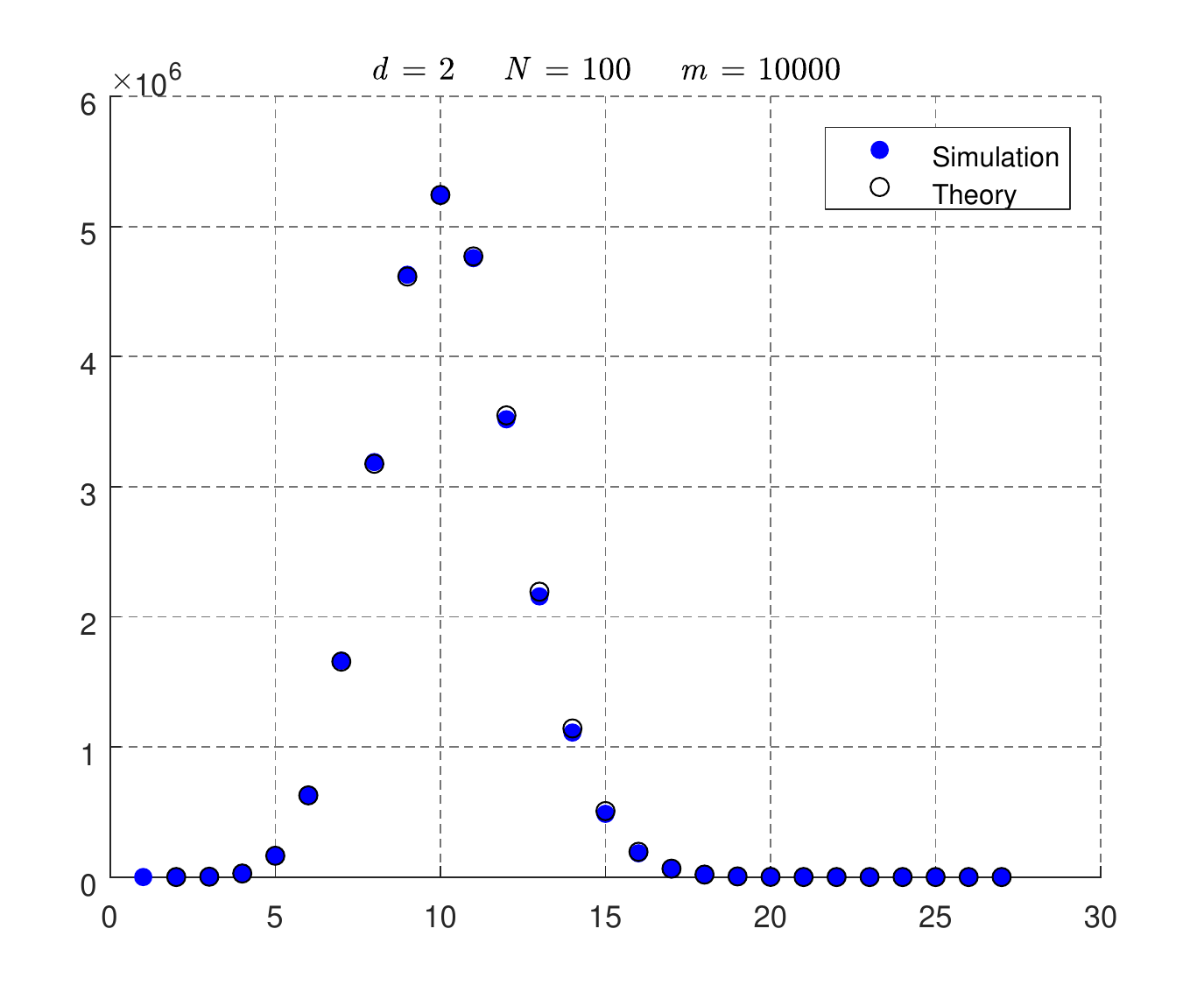}
\includegraphics[width=.45\textwidth]{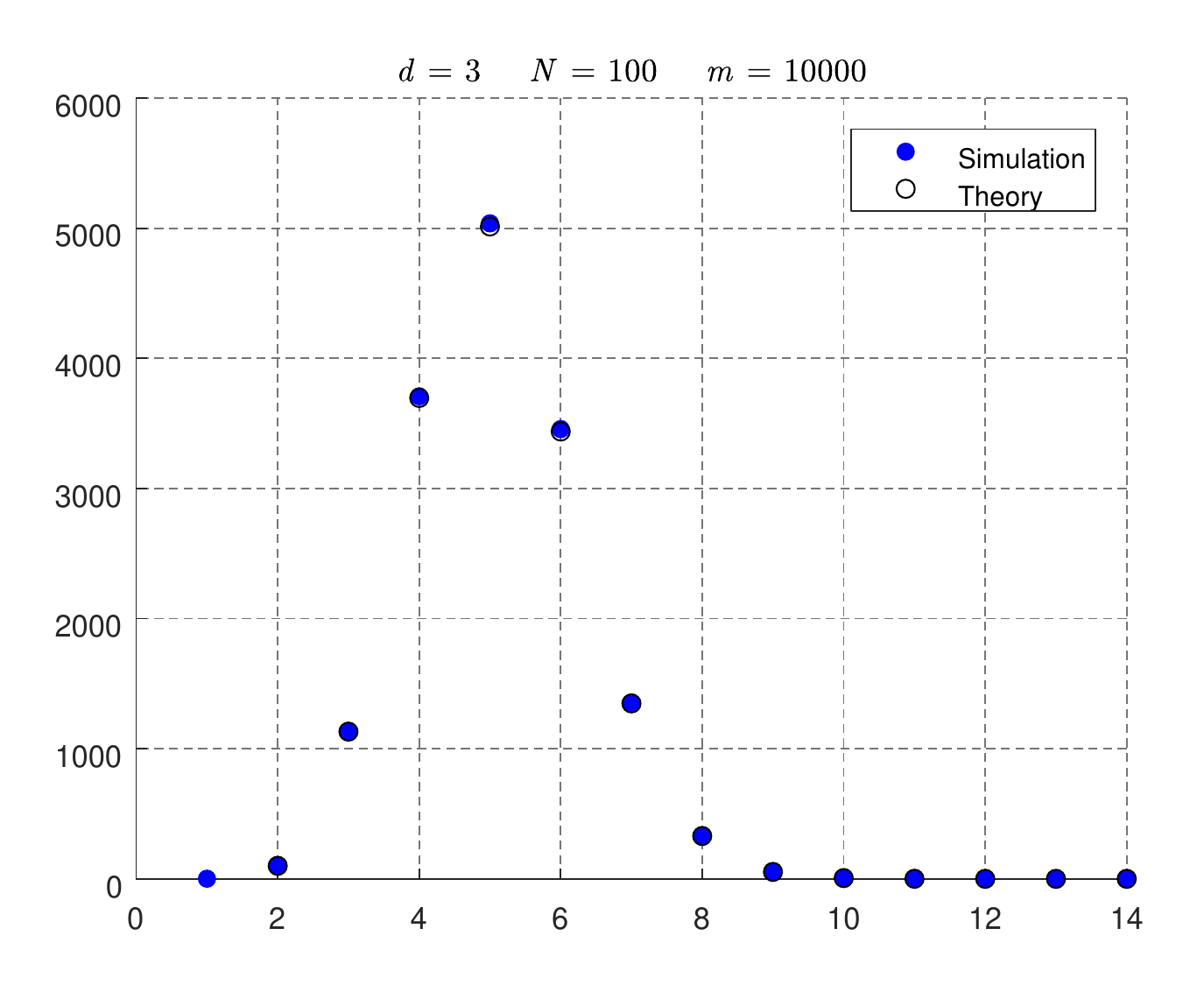}
\includegraphics[width=.45\textwidth]{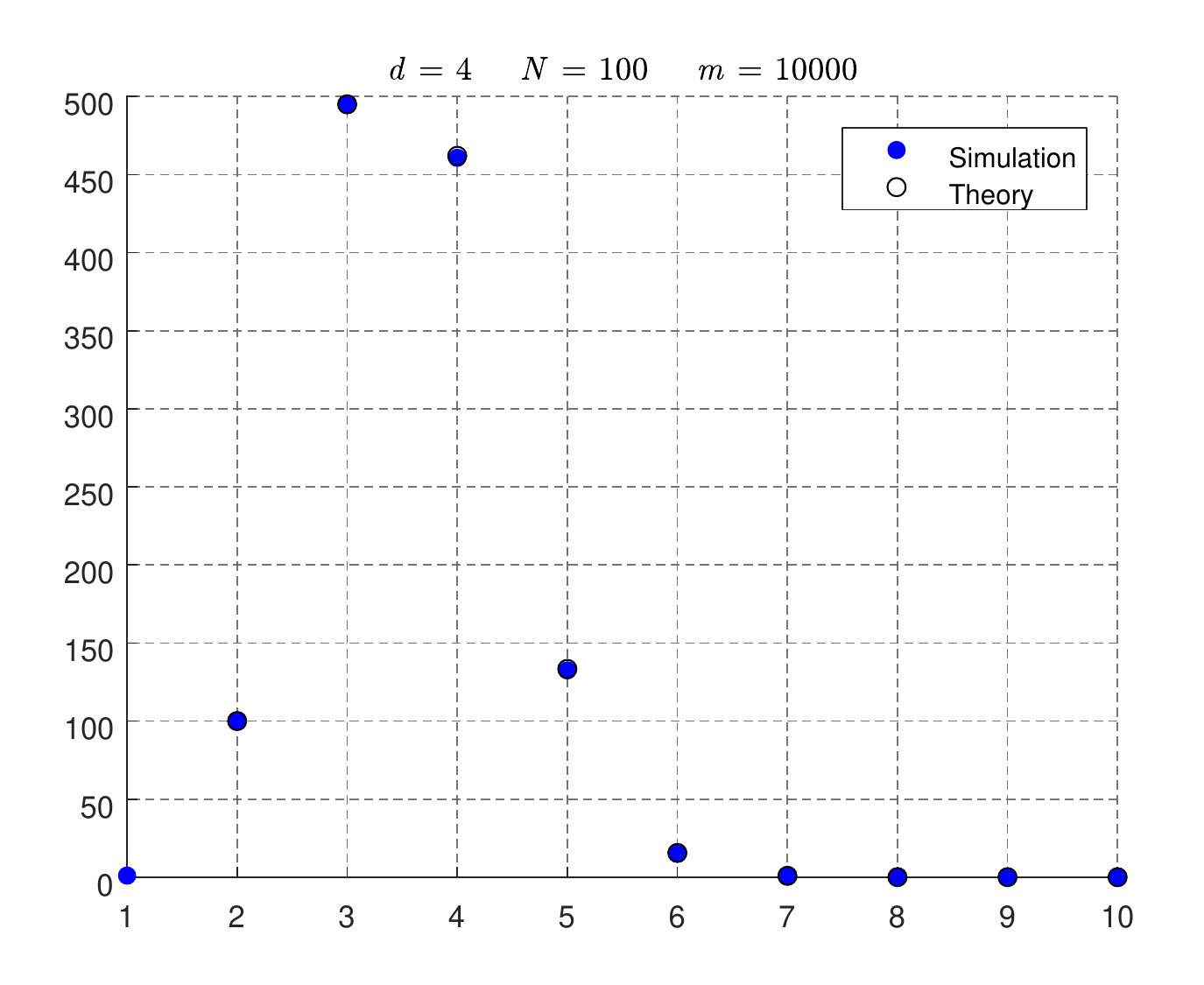}
\includegraphics[width=.45\textwidth]{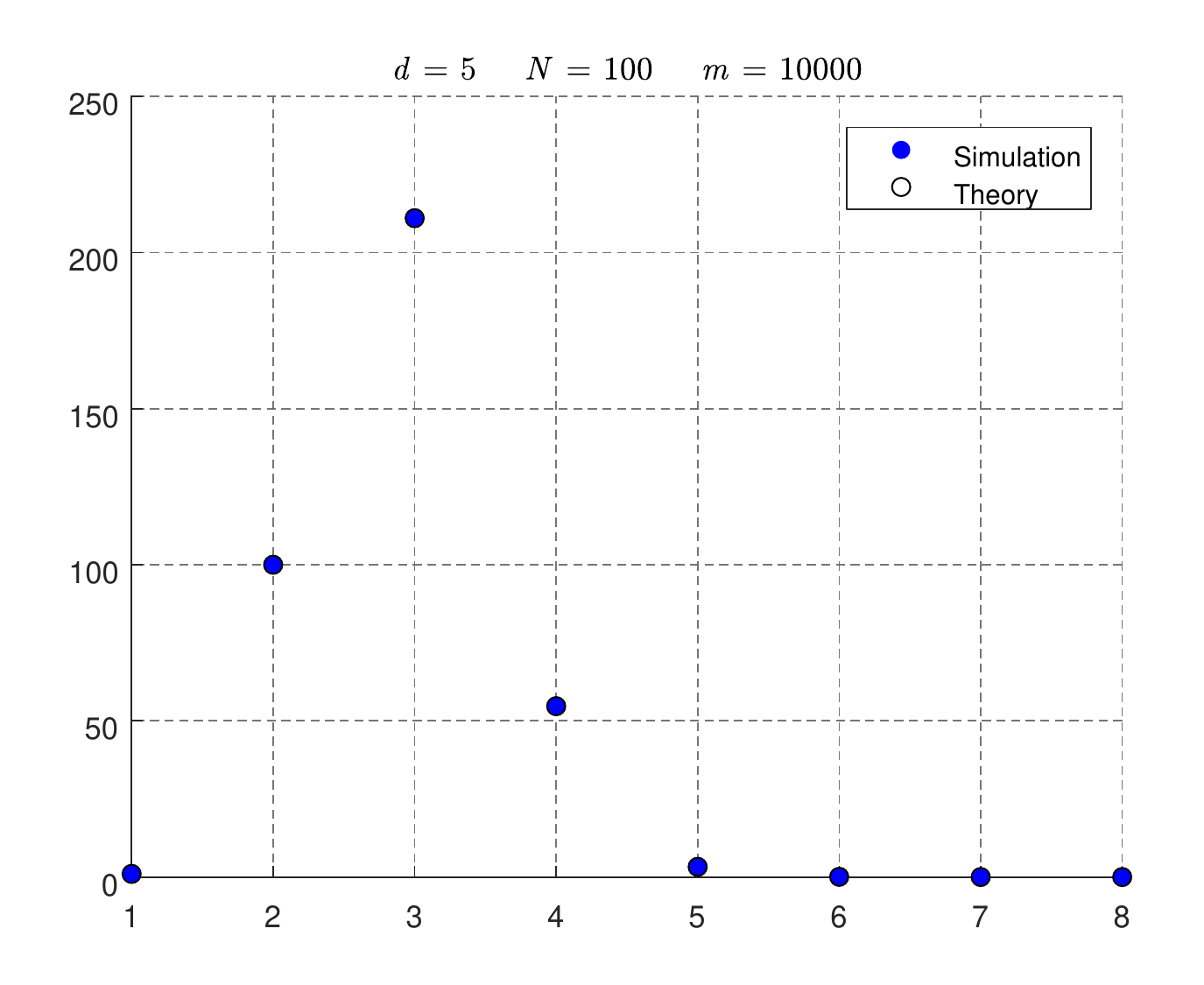}
\end{center}
\caption{Plots of the average number of $k$-chains as a function of $k$ for $m = 10,000$ simulated sprinklings of $N = 100$ points in Minkowski space of various dimensionalities, and the mean number of $k$-chains computed using the distribution (\ref{chainsO}) with corrected densities.}
\label{fig:distsO}
\end{figure}

We should emphasize the fact that the densities $\rho_i$ are not all equal to each other. In the previous literature, calculations were carried out assuming that $\rho_i = \rho_0$ (this includes our own work \cite{paths}, as well as references in which the $\rho_i = \rho_0$ assumption was hidden by the use of units in which $\rho = 1$ \cite{Meyer}), in which case instead of Eq.\ \ref{chainsO} one gets the chain-length distribution 
\beq
   \left<c_k\right>_{\rho_i=\rho_0} = N^{k-1}
   \left(\frac{\Gamma(d+1)}{2}\right)^{\!k-2}
   \frac{\Gamma(d/2+1)\,\Gamma(d)}{\Gamma((k-1)d/2+1)\,\Gamma({kd}/{2})}\;.
   \label{chainsM}
\eeq
(We should note that our $k$-chains are called $(k+1)$-chains in Ref.\ \cite{Meyer}, and our $d$ is the full spacetime dimension.) The difference between the values obtained using the two distributions (\ref{chainsO}) and (\ref{chainsM}) can be seen by comparing Figs.\ \ref{fig:distsM} and \ref{fig:distsO}. The former shows the results of $m = 10,000$ simulated random causal sets with $N = 100$ in Minkowski space of various dimensionalities together with the theoretical distributions from Eq.\ \ref{chainsM} for $N = 100$, while in the latter figure the distribution in Eq.\ \ref{chainsO} is used with the same set of random causal sets. It appears that, at least for dimensions 2-5, the $\rho_i = \rho_0$ distribution predicts more chains than are actually found, in particular for higher values of $k$, while with the corrected values for $\rho_i$ are used the fit is so good that in most cases it's difficult to even see the theory data points as they lie directly underneath the simulation data points. The effect is less dramatic in higher dimensions, but this is due to the fact that an $N$-point manifoldlike causal set will have shorter chains if it's embedded in higher dimensions. (We should be able to see the same dramatic differences in higher dimensions if we used enough points for the peak to be in the same place it is in the 2D case; however, this is very demanding computationally as the peak is around $N^{1/d}$.)

From the analytical expression for the mean number of chains of length $k$ we can now derive the mean chain length and the width of the chain-length distribution. The results of these calculations will be discussed in Section \ref{Manifoldlikeness}.

\section{Path Length Distributions in Minkowski Space}

In this section we derive the path-length distribution for a causal set uniformly embedded in Minkowski spacetime. We had obtained an expression for this distribution in two dimensions in Ref.\ \cite{paths}, but here we apply the correction introduced in the previous section to the expressions for the point densities and an extra modification to the path length distribution, and extend the previous results to an arbitrary number $d$ of dimensions. The setup for the problem is similar to the one used in Ref.\ \cite{paths}. We start from an expression for the probability of finding a $k$-path through a set of locations in Minkowski space analogous to the ones in Eqs.\ \ref{P2}--\ref{Pk} for the probability of finding a $k$-chain through that set of locations. The difference is that in the case of paths the probabilities must include factors expressing the requirement that none of the Alexandrov sets $A_{i,i+1}$ contain any other sprinkled points and therefore their union, of volume $V = \sum_i V_{i,i+1}$, be empty,
\beq
   \dd P_{(k)}(x_1,...,x_{k-1})
   = (\rho_0\,\dd V_1)\,(\rho_1\dd V_2)\,\cdots\,(\rho_{k-2}\dd V_{k-1})
   \left(1-\frac{\sum_{i=0}^{k-1}V_{i,i+1}}{V_0}\right)^{\!N-k+1}\;,
   \label{Pkp}
\eeq
for all $2 \le k \le N+1$ (while now for $k = 1$ the probability vanishes), where the binomial distribution has been used for the probability that the Alexandrov sets are empty, so now the integrated mean number of paths is given by
\beq
   \left<n_k\right> = \rho_0\int_{A_0}{\rm d}^dx_1\;\rho_1\int_{A_1}{\rm d}^dx_2
   \cdots
   \;\rho_{k-2}\int_{A_{k-2}}{\rm d}^dx_{k-1}\left(1-\frac{\sum_{i=0}^{k-1}
   V_{i,i+1}}{V_0}\right)^{\!N-k+1},
\eeq
where the densities $\rho_i$ were defined in Sec.\ \ref{chainlengths}, and the exponent of the argument in the integral is $N-k+1$, as this is the number of remaining points, the ones not in the path. It is important to note that as the length of the path increases the importance of this correction increases. Using a polynomial expansion for the argument we find
\bea
   && \left<n_k\right> = \frac{N!}{(N-k+1)!}\frac{1}{V_0^{k-1}}
   \sum_{i_1 = 0}^{N-k+1} {N-k+1\choose i_1} \left(\frac{-1}{V_0}\right)^{i_1}
   \sum_{i_2=0}^{i_1}{i_1\choose i_2}\cdots
   \sum_{i_k=0}^{i_{k-1}}{i_{k-1}\choose i_k}\times\notag\\
   && \kern30pt\times\int_{A_0}{\rm d}^dx_1\, V_{0,1}^{i_1-i_2}
   \int_{A_1}{\rm d}^dx_2\, V_{1,2}^{i_2-i_3} \cdots
   \int_{A_{k-2}}{\rm d}^dx_{k-1}\, V_{k-2,k-1}^{i_{k-1}-i_k}\, V_{k-1,k}^{i_k}\;.
\eea
These integrals can be explicitly calculated, and the result is
\bea
   && \left<n_k\right> = \frac{N!}{(N-k+1)!}
   \left(\frac{\Gamma(d+1)}{2\Gamma(d/2)}\right)^{\!k-1} \times\notag \\
   && \kern30pt\times\ \sum_{i=0}^{N-k+1}{N-k+1\choose i}\,
   \frac{(-1)^i\,\Gamma(i+1)}{\Gamma((i+k)d/2)\,\Gamma(1+(k-1+i)d/2)}
   \,f_{i,k}\;, \label{nk}
\eea
where the $f_{i,k}$ are defined by
\beq
   f_{i,1} = \frac{\Gamma((i+1)d/2)\Gamma(1+id/2)}{\Gamma(i+1)}\;. \label{fi1}
\eeq
for $k = 1$, $0 \le i \le N$, and
\beq
   f_{i,k}
   = \sum_{j=0}^i\frac{\Gamma((i-j+1)d/2)\Gamma(1+(i-j)d/2)}{\Gamma(i-j+1)}
   \,f_{j,k-1}\;, \label{fik}
\eeq
for $2 \le k \le N+1$, $0 \le i \le N-k+1$. We will prove the result (\ref{nk}) by induction. Let us assume that the result is correct for $\left<n_k\right>$, consider the expression for $\left<n_{k+1}\right>$, and show that we can write the latter in the same form. Written for $\left<n_{k+1}\right>$, Eq.\ (\ref{nk}) becomes
\bea
   &&\left<n_{k+1}\right> = \frac{N!}{(N-k)!}\frac{1}{V_0^k}
   \sum_{i_1=0}^{N-k}{N-k\choose i_1}\left(\frac{-1}{V_0}\right)^{i_1}
   \sum_{i_2=0}^{i_1}{i_1\choose i_2}\cdots\sum_{i_k=0}^{i_k}
   {i_k\choose i_{k+1}}\times\notag\\
   &&\kern40pt \times\int_{A_0}{\rm d}^dx_1\ V_{0,1}^{i_1-i_2}
   \int_{A_1}{\rm d}^dx_2\, V_{1,2}^{i_2-i_3}\cdots
   \int_{A_{k-1}}{\rm d}^dx_k\ V_{k-1,k}^{i_k-i_{k+1}}V_{k,k+1}^{i_{k+1}}\;.
\eea
We see that this equation has one more integral, resulting in one more parameter and hence one more sum. If we use the result in Eq.\ (\ref{nk}), our induction hypothesis, to evaluate the first $k-1$ integrals and the corresponding sums, some simple algebra gives
\bea
\label{induct}
   &&\left<n_{k+1}\right> = \frac{N!}{(N-k)!}\frac{1}{V_0^k}
   \left(\frac{\Gamma(d+1)}{2\Gamma(\tfrac{d}{2})}\right)^{k-1}
   \sum_{i_1=0}^{N-k}{N-k\choose i_1}\left(\frac{-1}{V_0}\right)^{i_1}
   \times\notag\\
   &&\times\sum_{i_2=0}^{i_1}{i_1\choose i_2}
   \frac{\Gamma(i_2+1)}{\Gamma((i_2+k)
   \tfrac{d}{2})\Gamma(1+(k-1+i_2)\tfrac{d}{2})}f_{i_2,k}
   \int_{A_0}{\rm d}^dx_1\ V_{0,1}^{i_1-i_2}V_{1,k+1}^{i_2+k-1}.
\eea
The integral can be simply calculated using null coordinates defined in terms of the Minkowski time $t$ and radial coordinate $r = (\sum_{i=1}^{d-1}x_i^2)^{1/2}$ as
\beq
	u = (t+r)/\sqrt{2}\;,\qquad v = (t-r)/\sqrt{2}\;,
\eeq
which for the last integral factor in the above equation gives
\bea
   &&\int_{A_0}{\rm d}^dx_1\ V_{0,1}^{i_1-i_2}V_{1,k+1}^{i_2+k-1} \\
   &&\kern0pt =V_0^{i_1+k}\frac{\Gamma(d+1)}{2\Gamma(\tfrac{d}{2})}
   \frac{\Gamma(\tfrac{d}{2}(i_2+k))\Gamma(1+\tfrac{d}{2}(k+i_2-1))}
   {\Gamma(\tfrac{d}{2}(i_1+k+1))\Gamma(1+\tfrac{d}{2}(k+i_1))}
   \,\Gamma(1+\tfrac{d}{2}(i_1-i_2))\,\Gamma(\tfrac{d}{2}(i_1-i_2+1))\;.\notag
\eea
After plugging this result back into Eq.\ \eqref{induct} and some simple algebra we obtain
\bea
   &&\left<n_{k+1}\right> = \frac{N!}{(N-k)!}
   \left(\frac{\Gamma(d+1)}{2\Gamma(d/2)}\right)^{\!k}
   \,\sum_{i_1=0}^{N-k}{N-k\choose i_1}
   \frac{\Gamma(i_1+1)}{\Gamma(\tfrac{d}{2}(i_1+k+1))\Gamma(1+\tfrac{d}{2}(k+i_1))}
   \times\notag\\
   &&\times\underbrace{\sum_{i_2=0}^{i_1}\frac{\Gamma(\tfrac{d}{2}(i_1-i_2+1))
   \Gamma(1+\tfrac{d}{2}(i_1-i_2))}{\Gamma(i_1-i_2+1)}f_{i_2,k}}_{f_{i_1,k+1}}\;,
\eea
where we have used the recursion relation defined in Eq.\ (\ref{fik}). The above equation is exactly the equation for $\left<n_k\right>$ with $k$ replaced with $k+1$.

This result is in complete agreement with the 2-dimensional case obtained in Ref.\ \cite{paths}, with the exception of the small corrections mentioned above. It is also interesting to note that the term with $i=0$ in the sum corresponds to the number of chains, which was discussed in Sec.\ \ref{chainlengths}. We now turn our attention to applications of these distributions.

\begin{figure}[htp]
\begin{center}
\includegraphics[width=.45\textwidth]{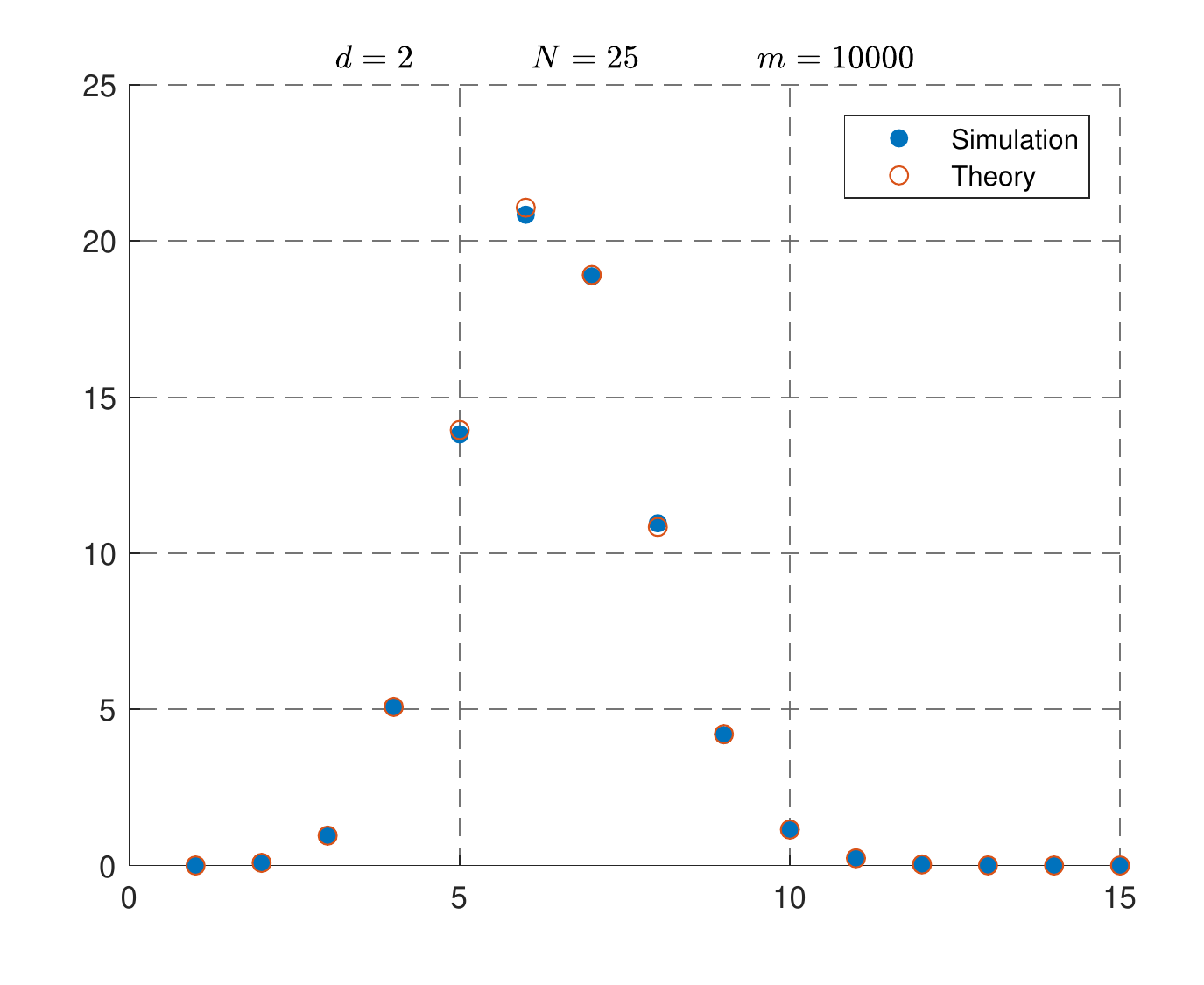}
\includegraphics[width=.45\textwidth]{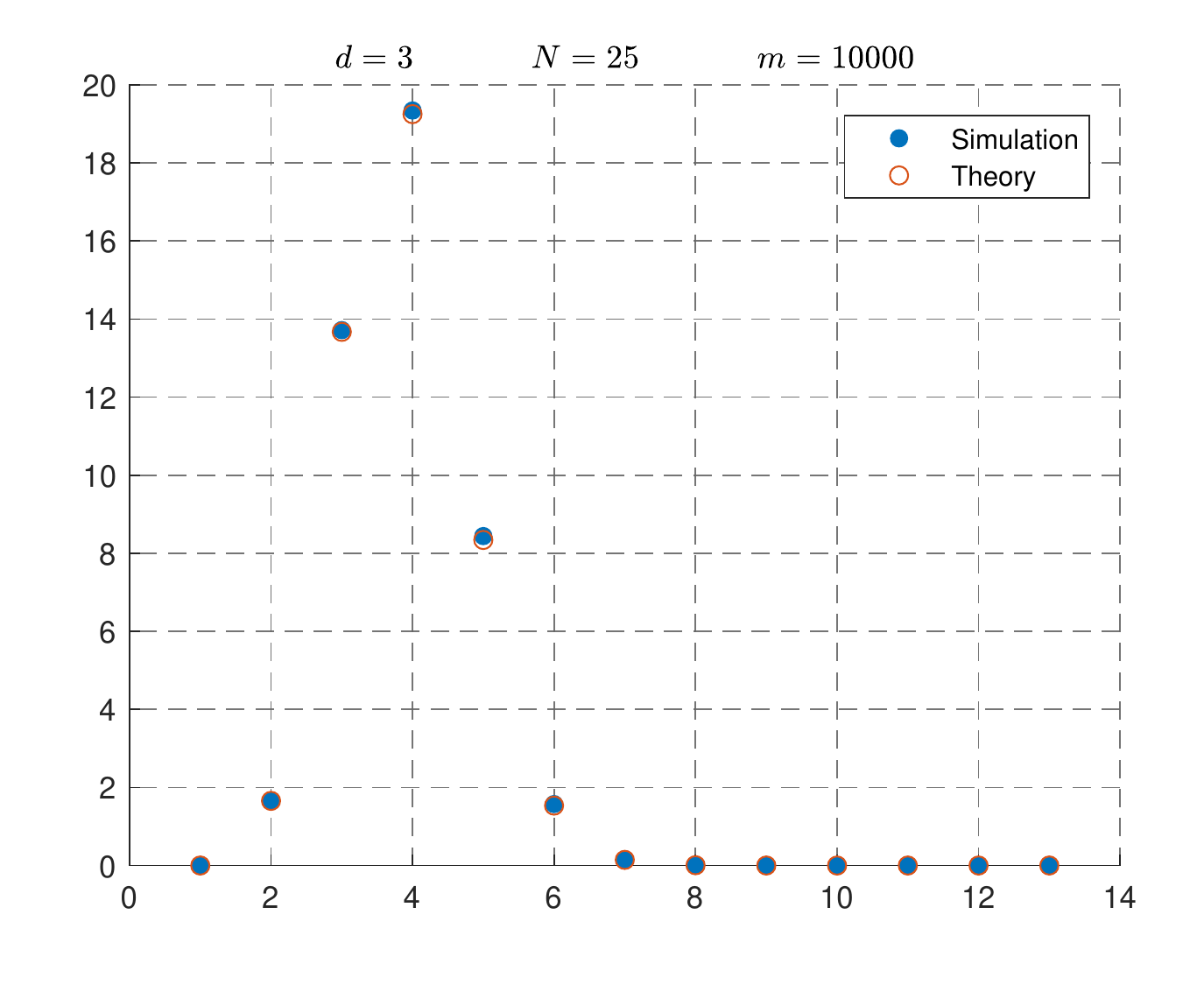}
\includegraphics[width=.45\textwidth]{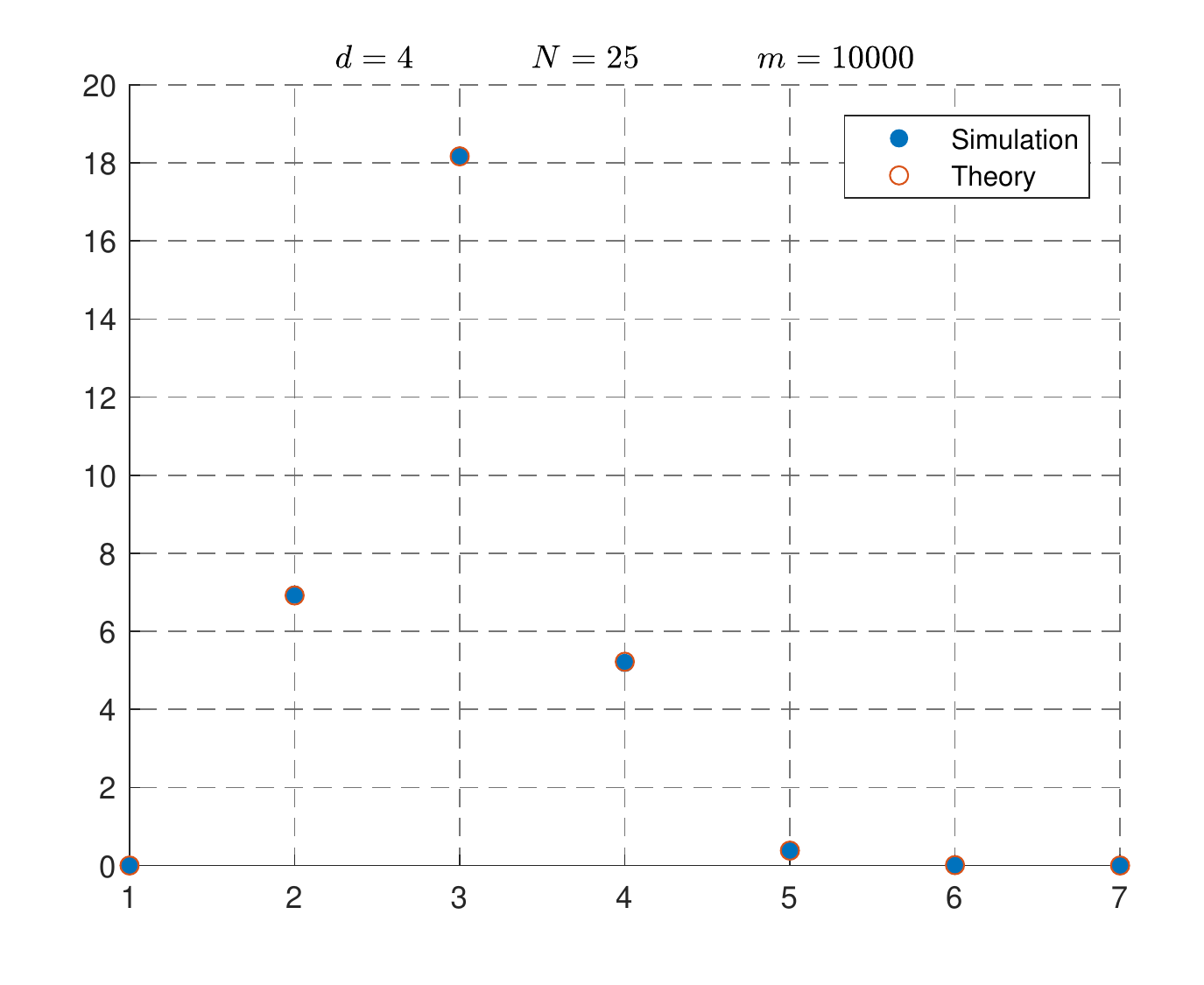}
\includegraphics[width=.45\textwidth]{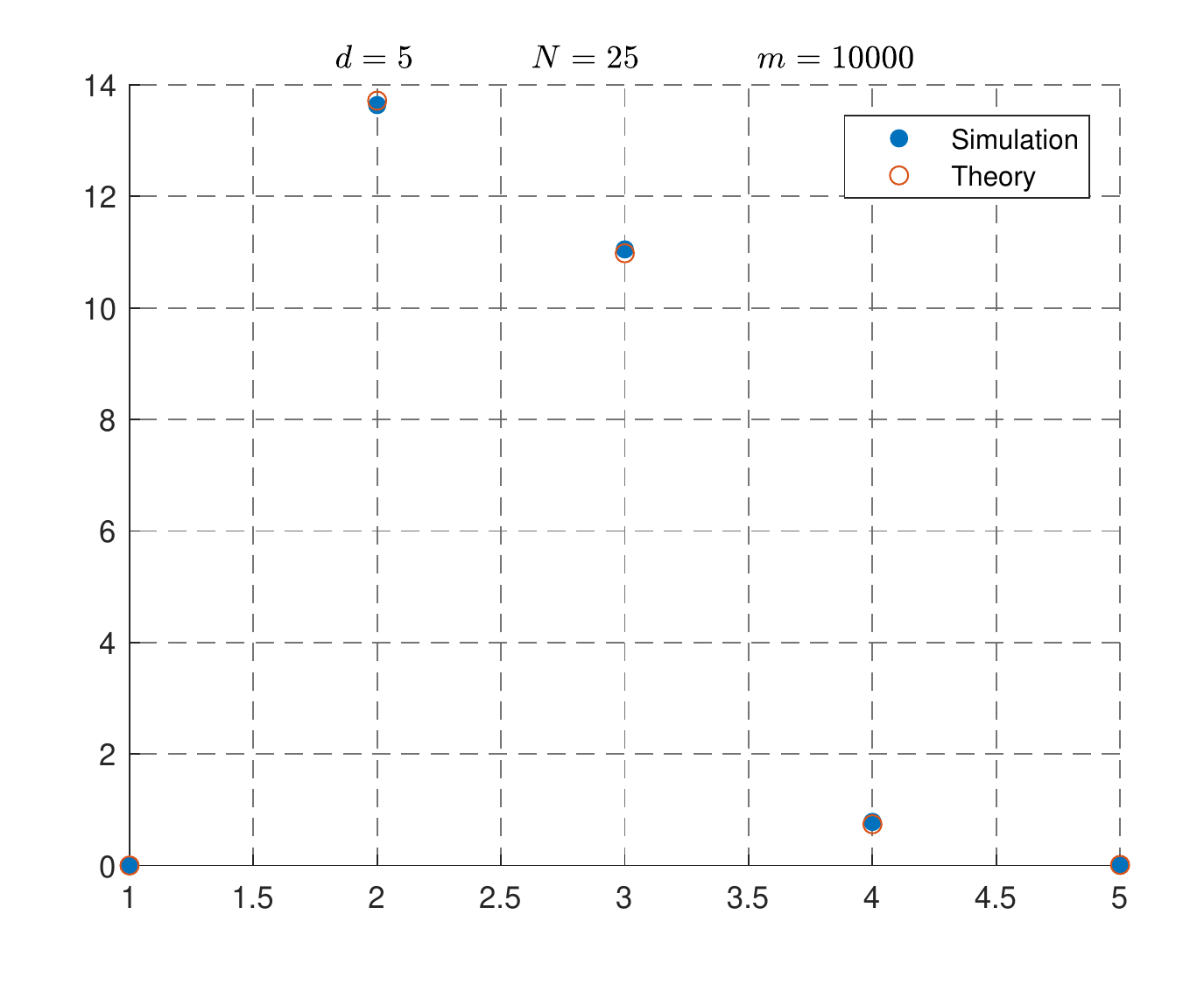}
\end{center}
\caption{Plots of the average number of $k$-paths as a function of $k$ for $m = 10,000$ simulated sprinklings of $N = 25$ points in Minkowski space of various dimensionalities, and the mean number of $k$-paths computed from the path-length distribution (\ref{nk}) that uses the corrected densities.}
\label{fig:pathscompare}
\end{figure}

\section{Example: Manifoldlikeness}\label{Manifoldlikeness}

A standing question in causal set theory is how to determine whether a causal set $\mathcal C$ is manifoldlike in the sense that it is well approximated by a continuum at large scales. In other words, is there a mapping of the causal set to a manifold $\mathcal M$ such that the causal relations are preserved and the points are distributed in a way that accurately probes the geometry of the continuum? There has been some progress on general aspects of this question, with results showing that if a causal set is manifoldlike, the manifold it reproduces is ``approximately unique" \cite{BTN,closeness}. Our goal here is to address more specific aspects.

\subsection{Paths}

\begin{figure}
\centering
\begin{minipage}{.45\textwidth}
\includegraphics[width=\textwidth]{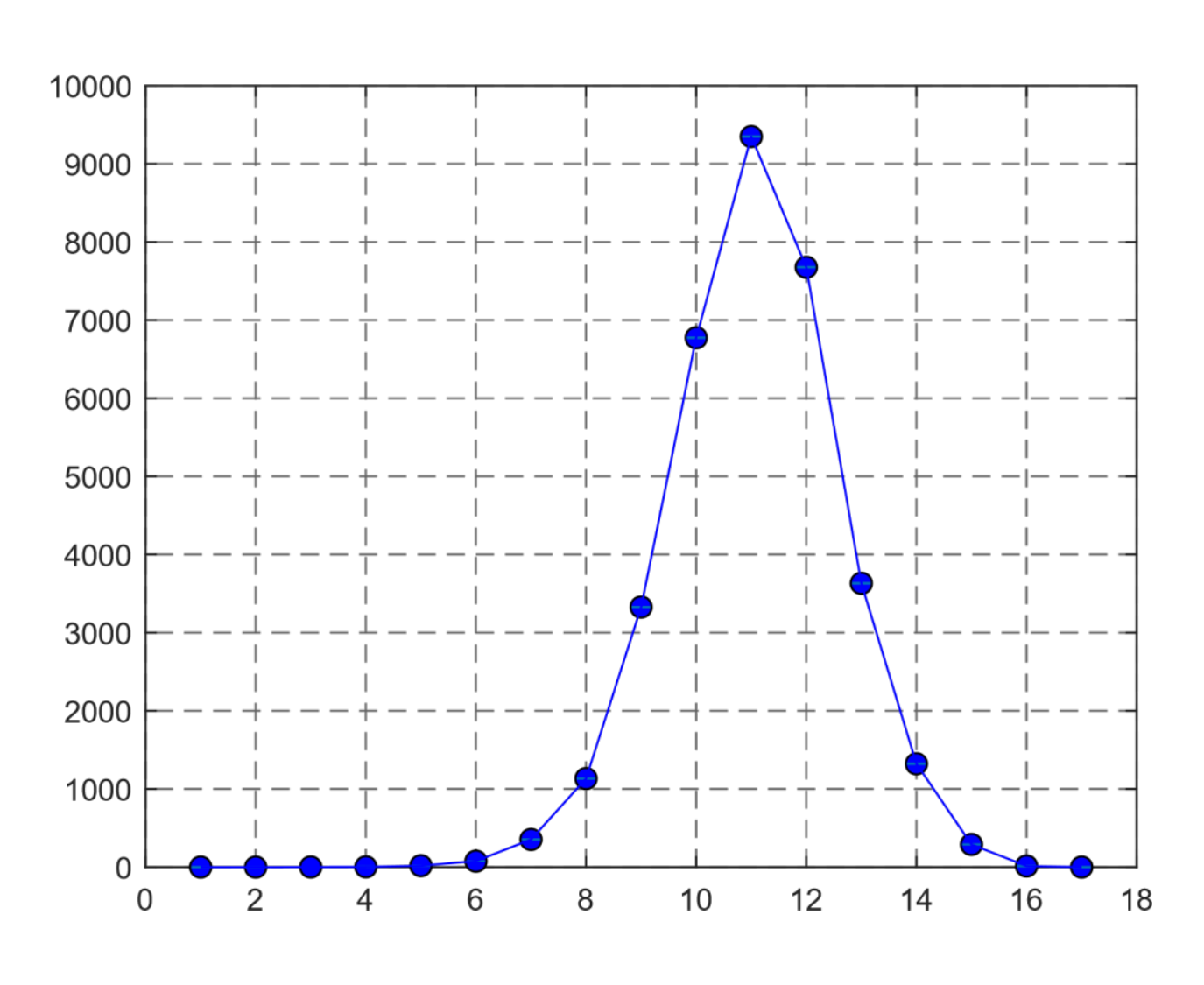}
\end{minipage}
\begin{minipage}{.45\textwidth}
\centering
\includegraphics[width=\textwidth]{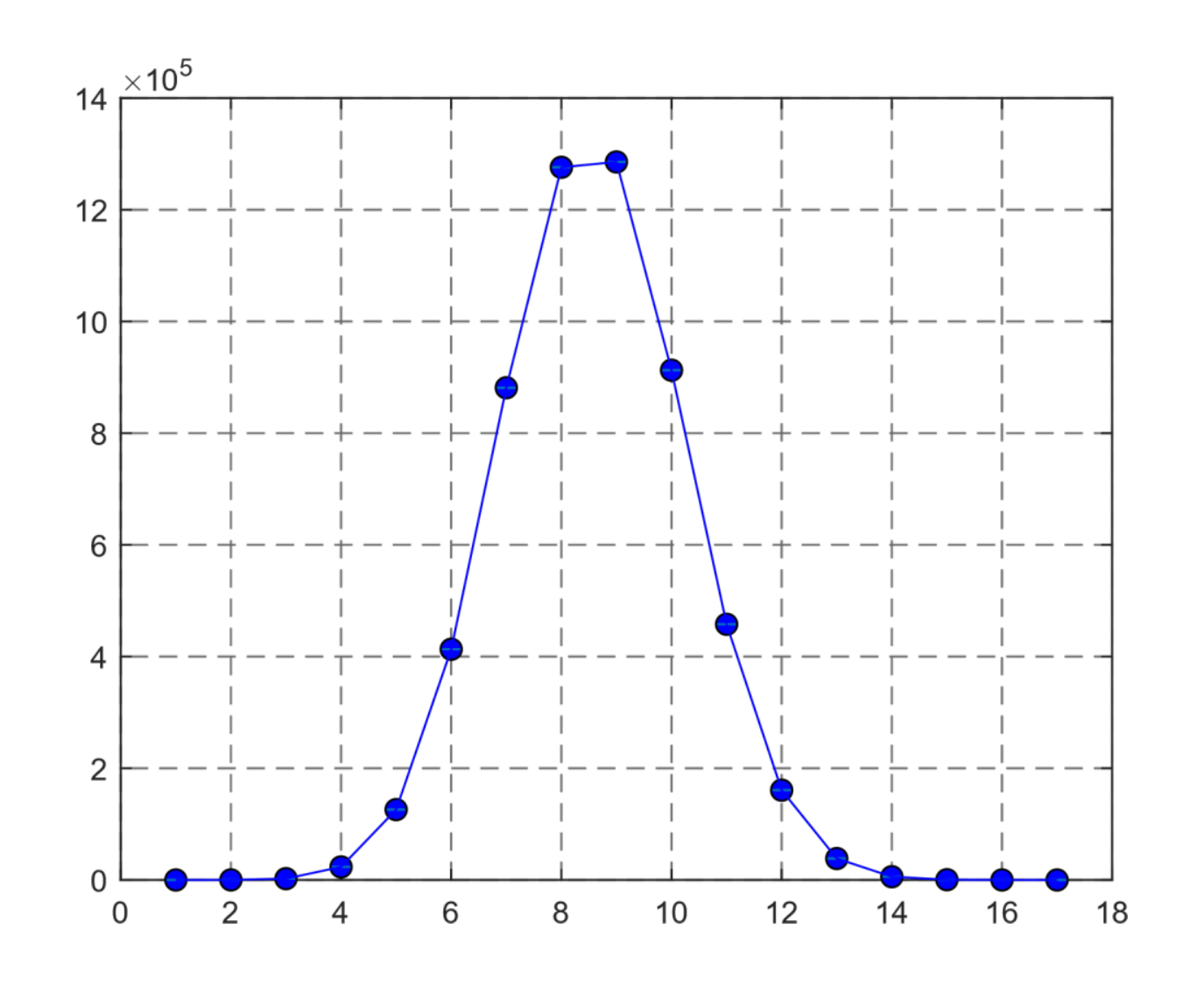}
\end{minipage}
\caption{Plots of the path and chain length distributions $n_k$ and $c_k$ (left and right, respectively) of a single causal set of size $N=100$ embedded in 2-dimensional Minkowski space.}
\label{fig:paths}
\end{figure}

In Ref.\ \cite{paths} we argued that a causal set's path length distribution could be used to determine if the set was manifoldlike, at least in Minkowski space, by comparing it to the theoretical one. We will briefly recall this argument here. The path length distributions of manifoldlike causal sets all share a set of characteristics which distinguish them from at least most (perhaps all) non-manifoldlike distributions. Manifoldlike causal sets have strongly peaked, Gaussian-looking path length distributions, as seen from the left part of Fig.\ \ref{fig:paths}, with mean peak position, height and width (and size of fluctuations around those values) depending on $d$ and $N$ in ways that can be obtained from simulations or calculated analytically. Most non-manifoldlike causal sets, however, have path length distributions which are either qualitatively different or do not fit the known dependence on $d$ and $N$.

\begin{figure}
\centering
\begin{minipage}{.45\textwidth}
\centering
\includegraphics[width=\textwidth]{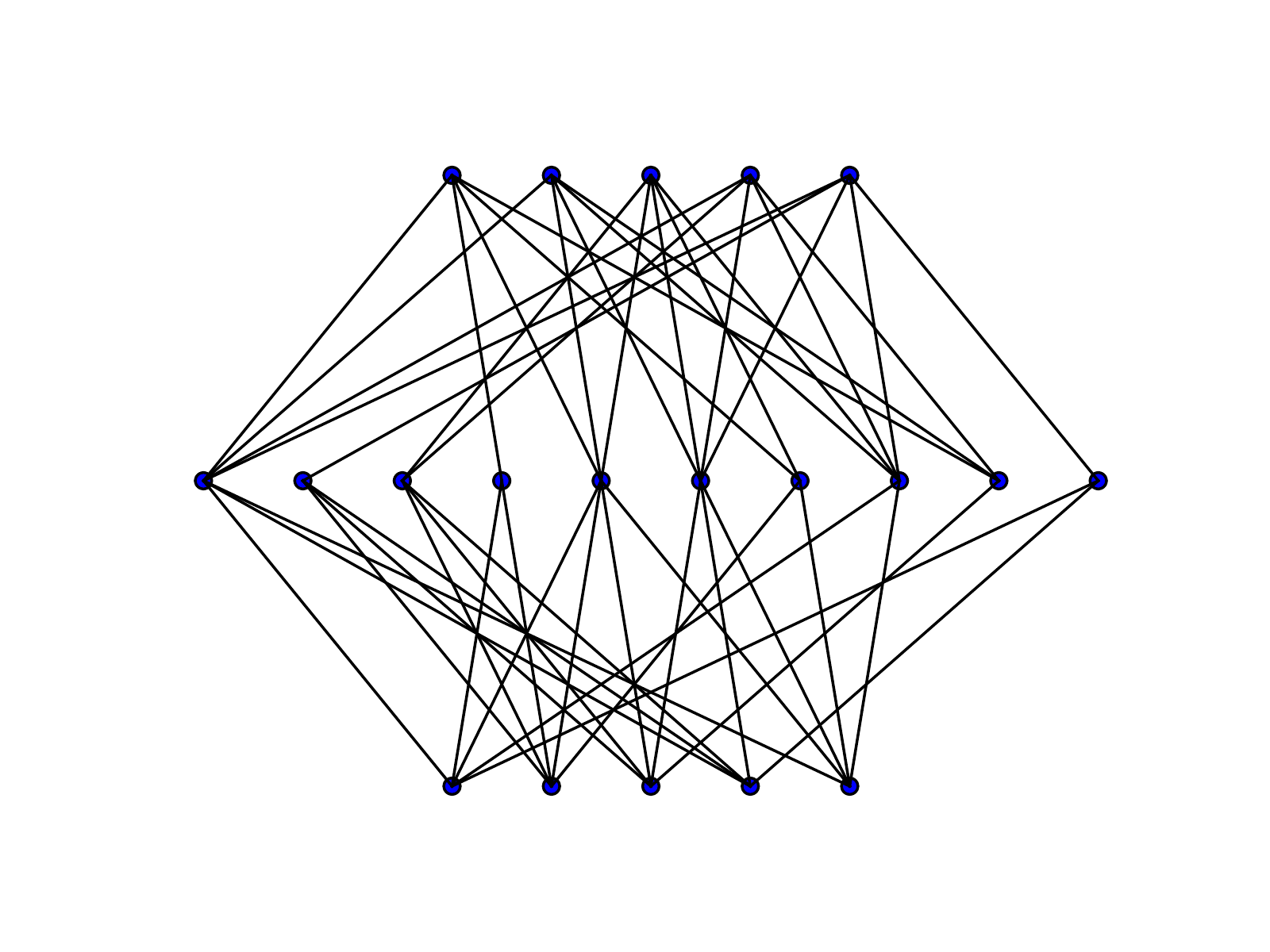}
\end{minipage}
\begin{minipage}{.45\textwidth}
\centering
\includegraphics[width=\textwidth]{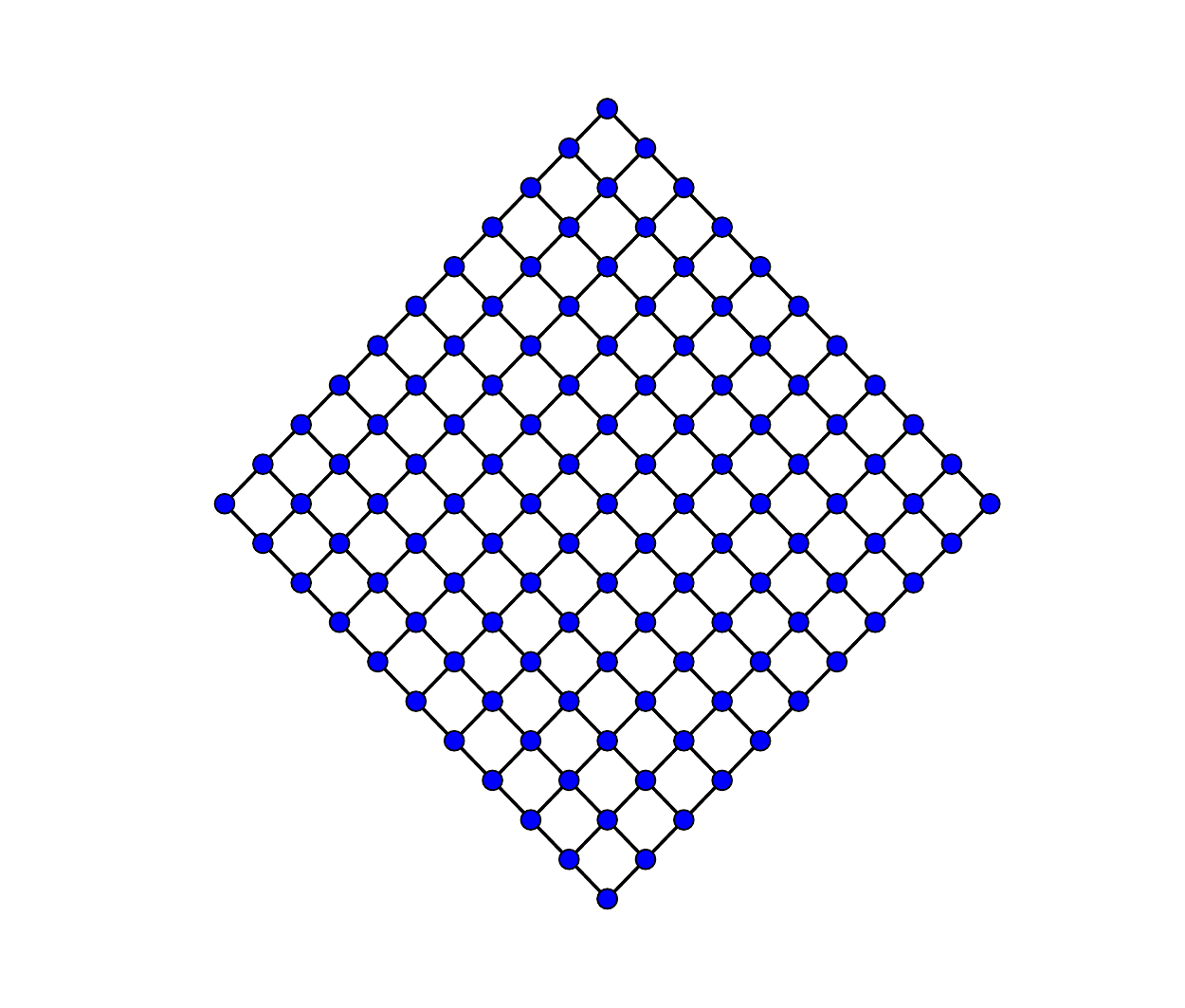}
\end{minipage}
\caption{Left: A Kleitman-Rothschild causal set. Right: An $11\times11$ square lattice.}
\label{fig:KR_2Dlat}
\end{figure}

While there are many different types of non-manifoldlike causal sets, we will focus on two of the most obvious ones. Most very large causal sets (the fraction goes to 1 as $N\to\infty$) are of the Kleitman-Rothschild type \cite{KR}. These causal sets are in three layers, with a fourth of the points being in the top layer, a fourth in the bottom layer, and half in the middle layer; each element in the bottom and top layers is linked to half of the elements in the middle layer, as seen on the left of Fig.\ \ref{fig:KR_2Dlat}. These causal sets are obviously not manifoldlike as their height does not grow with size, and they are thus unable to encode all of the geometric information contained in a manifold, and their path length distribution is proportional to a Kronecker delta at $k = 2$.

The other non-manifoldlike causal sets we will consider are regular lattices. In particular, we will look at 2- and 3-dimensional cases in detail because they can be set up differently, but we expect the higher-dimensional cases to give similar results. Regular lattices are probably the most obvious way to discretize a geometry. However, they do not work for Lorentzian geometries in the sense that each one picks a preferred frame and is therefore not Lorentz-invariant. A 2D regular lattice can be set up for example with its lattice sites placed at evenly spaced locations along null lines, as in the right-hand part of Fig.\ \ref{fig:KR_2Dlat}. As the figure shows, each point that is not on the boundary is linked to exactly two other points along $45^\circ$ lines as measured from the vertical, and all links in this causal set are along one of those lines\footnote{Technically, to call these elements linked, we have to either draw the lines to be timelike rather than null, or modify our continuum interpretation of $a\prec b$ to correspond to causal relations as opposed to timelike ones.} (also, in this figure the lattice looks square, but that would not be the case in any boosted frame). This means that the path length distribution is again proportional to a Kronecker delta with, in the case of the $11\times 11$ lattice shown, $n_k \propto \delta_{k,20}$.

\begin{figure}
\centering
\begin{minipage}{.50\textwidth}
\centering
\includegraphics[width=\textwidth]{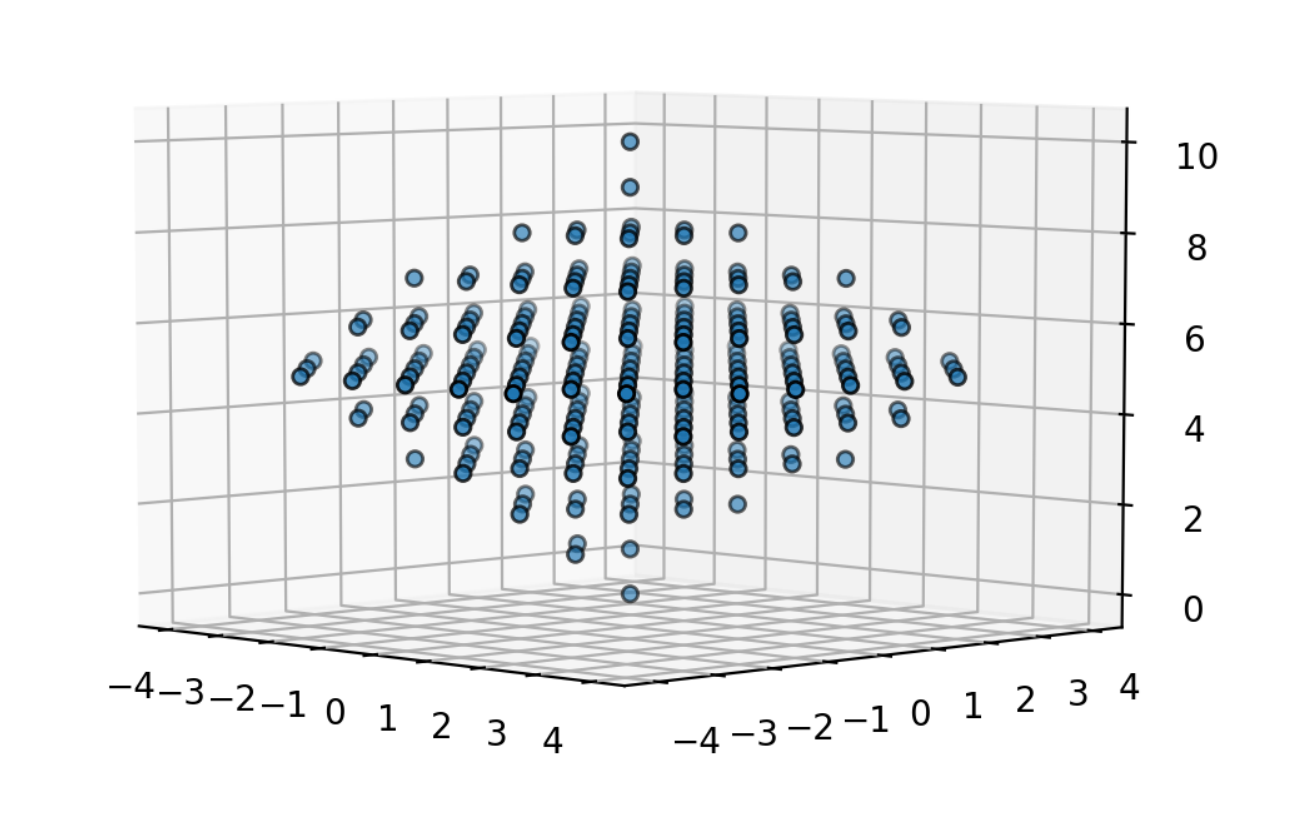}
\end{minipage}
\begin{minipage}{.40\textwidth}
\centering
\includegraphics[width=\textwidth]{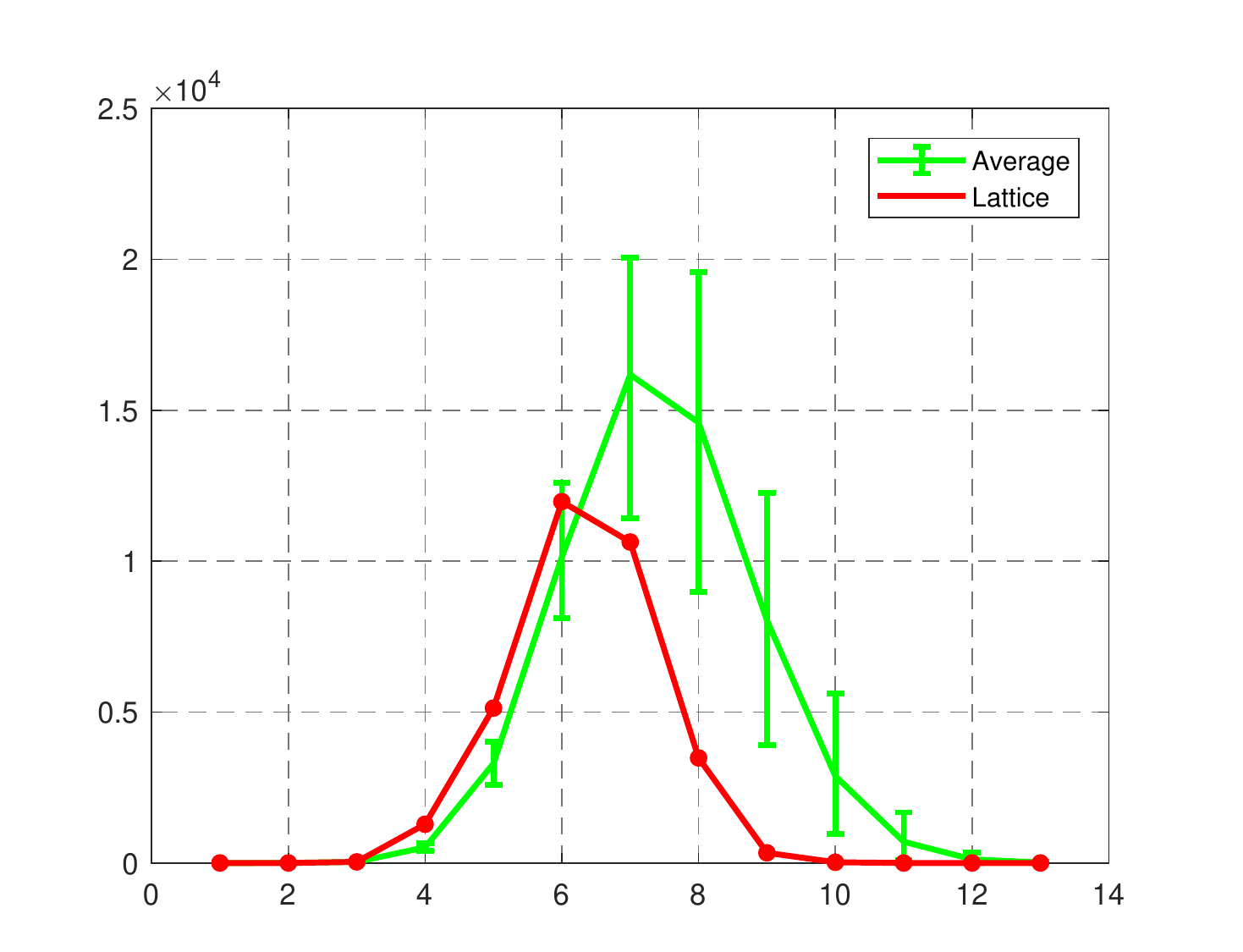}
\end{minipage}
\caption{Left: A 3-dimensional ``square" lattice (actually a cubic lattice in spacetime); the figure shows the portion of the lattice inside the Alexandrov set of the minimal and maximal points, which has $N = 235$ points. Right: The path length distribution for this square lattice, compared with the average of the corresponding distributions for 500 random $235$-point causal sets in 3D Minkowski space; the error bars indicate the range of values corresponding to the 50th percentile above and below the average.}
\label{fig:3Dlat}
\end{figure}

In the 3-dimensional case a regular lattice cannot have all points arranged along null lines, so we consider an example in which they are arranged along $t$, $x$ and $y$ coordinate lines of a Cartesian coordinate system. An example with again 11 layers in the time direction is shown on the left in \ref{fig:3Dlat}. This lattice has paths of varying lengths, the longest ones having $k = 10$. The whole path-length distribution is shown on the right in Fig.\ \ref{fig:3Dlat}, together with the corresponding average distribution for a sample of 20 causal sets randomly sprinkled in 3D Minkowski space. The two distributions differ in height, peak position and width. While we know that the height of the distributions for random causal sets in Minkowski space can fluctuate considerably, the peak position and width are subject to smaller fluctuations, and can be used to determine that the 3D regular lattice is not manifoldlike.

If we had just seen these distributions, we would be able to determine they were not manifoldlike. While there are non-manifoldlike causal sets which pass this criterion, at the very least (for large values of $N$) this program will allow us to eliminate most non-manifoldlike causal sets.

\subsection{Chains}

\begin{figure}[H]
\centering
\begin{minipage}{.45\textwidth}
\centering
\includegraphics[width = \textwidth]{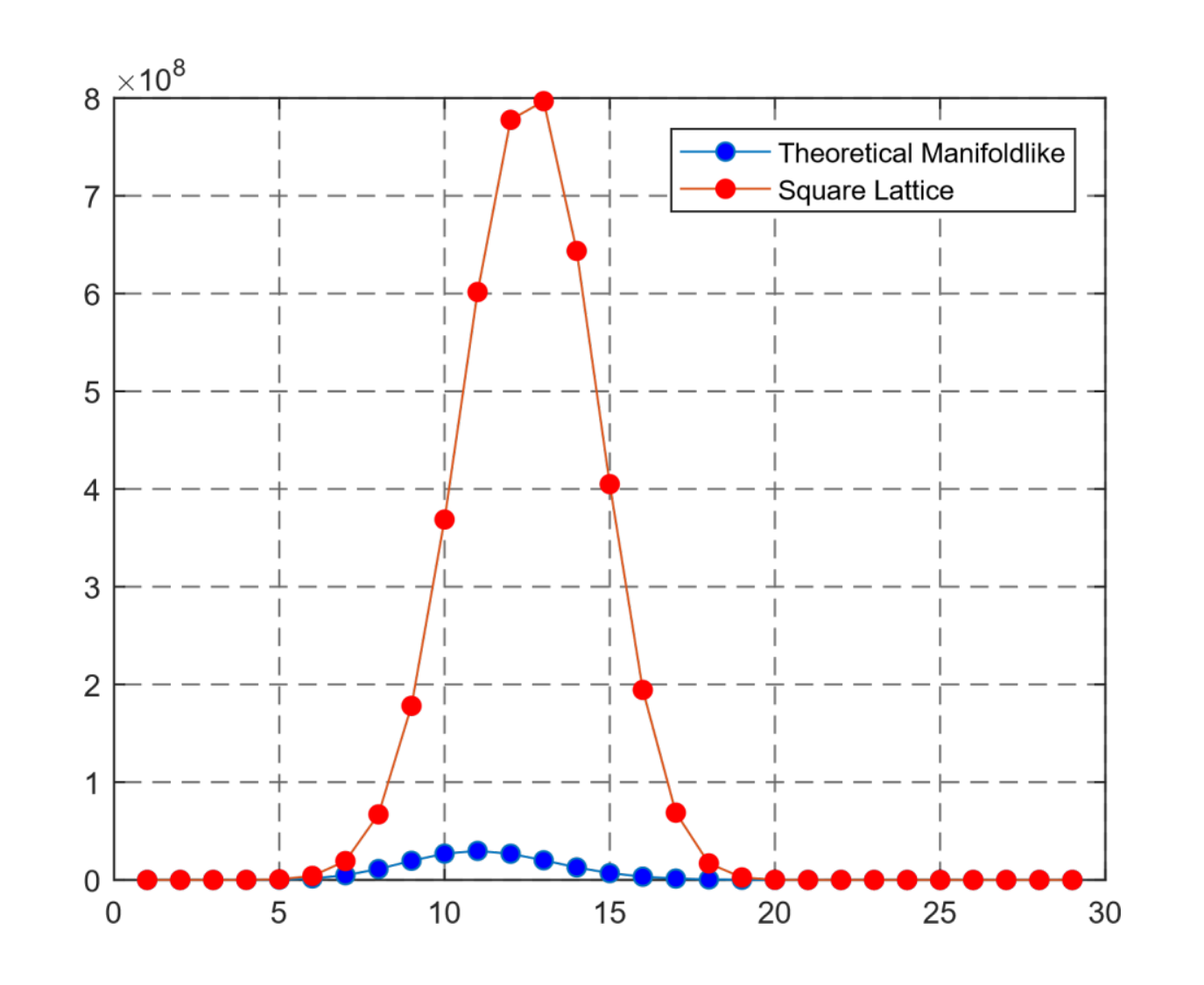}
\end{minipage}
\begin{minipage}{.45\textwidth}
\centering
\includegraphics[width = \textwidth]{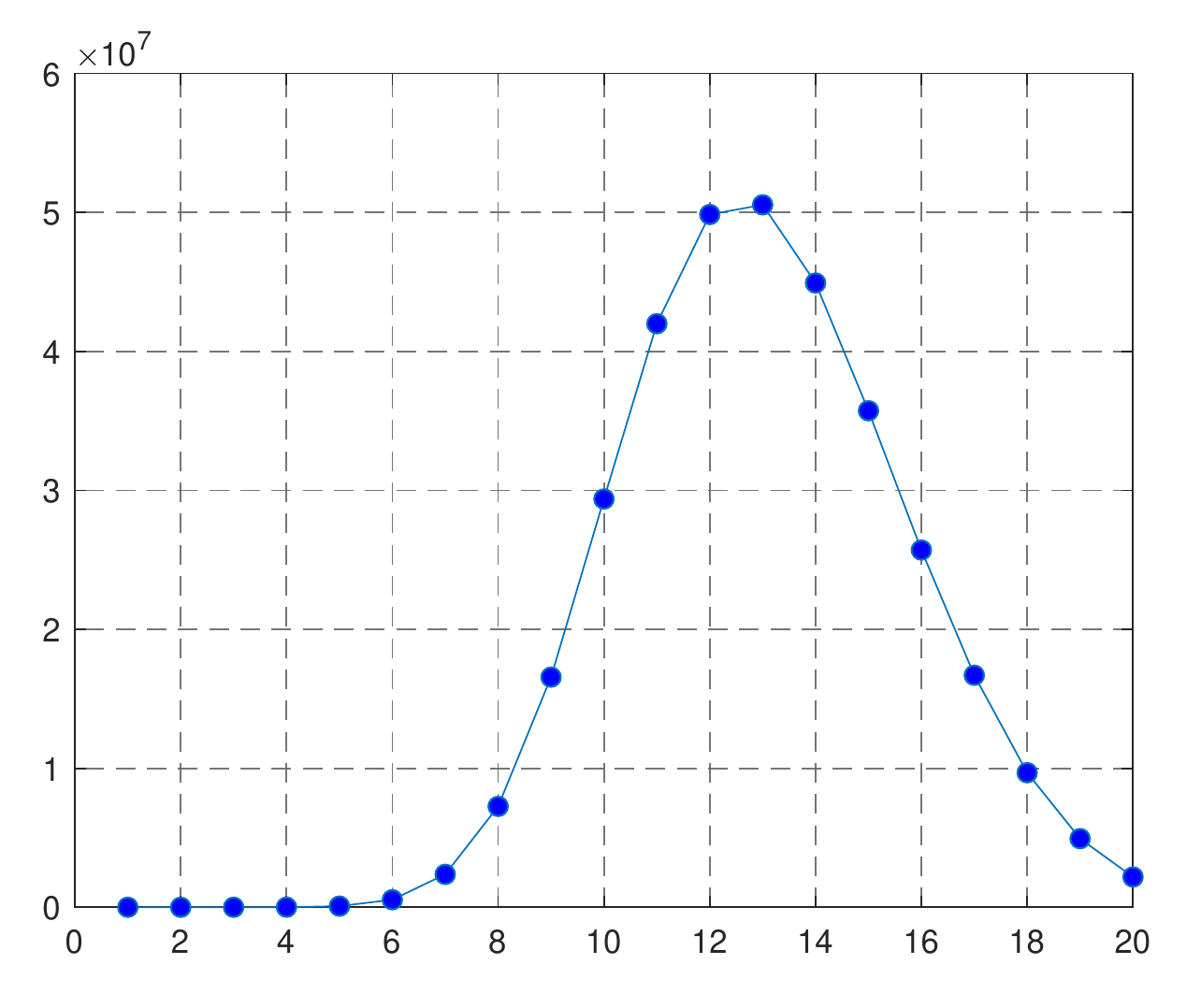}
\end{minipage}
\caption{\label{sigma} Left: Plot of the chain distributions of the 2-dimensional $11\times 11$ square lattice and the theoretical average chain distribution for causal sets of size $N = 119$ embedded in 2-dimensional Minkowski space; Right: Plot of the standard deviation of each $c_k$ for the 119-point 2D Minkowski causal sets on the left, as a function of $k$.}
\end{figure}

We now wish to use a causal set's chain distribution as a measure of manifoldlikeness; this may seem puzzling as we already have a measure using the path distribution; however, the expected number of chains is much easier to calculate, and we have a formula to find the expected number of chains of length $k$, whereas for paths we only have a recursion relation; furthermore, we can extend our calculation beyond Minkowski space using Riemann normal coordinates (possibly to an arbitrary order of $RV^{2/d}$ where $R$ is some curvature quantity) allowing us to broaden our measure of manifoldlikeness to causal sets embedded in curved manifolds. From the right of Fig.\ \ref{fig:paths} we can see that the chain length distribution of causal sets embedded in 2-dimensional Minkowski space has a shape similar to that of the path lengths; however, it is much taller, as there are many more chains than paths, and shifted to the left. Let's compare these distributions to those of the Kleitman-Rothschild type and our 2-dimensional square lattice. Neither distribution is as simple as its path counterpart, so let's add minimal and maximal elements for the Kleitman-Rothschild causal set; our chain length distribution then will have one chain of length one, $N$ chains of length two (all chain distributions share these features), and depending on the particular causal set some number of chains of length three and four. It's not quite the Kronecker delta from before, but it's obviously not the Gaussian like distributions of the manifoldlike causal sets. The square lattice, however, is a different story. From Fig.\ \ref{sigma}, we can see a number of things: the shape of the square lattice's chain distribution is similar to the theoretical distribution; however, it is much larger. This doesn't necessarily exclude it from being manifoldlike though, as individual distributions can have huge variations away from the average. What we can do is determine how many standard deviations away from the theoretical average the square lattice is. 

Computing $\sigma_k^2=\left<c_k^2\right>-\left<c_k\right>^2$ is difficult but not impossible; however, the number of terms we have to calculate increases rapidly with $k$, so let's restrict our attention to the simplest case. In an $N$-element causal set we always have $c_1 = 1$ and $c_2 = N$, so $k=3$ is the first case in which the calculation isn't trivial. This standard deviation can be calculated theoretically in the usual way, $\sigma^2_3 = \left<c_3^2\right>-\left<c_3\right>^2$. We know how to calculate $\left<c_3\right>$ and thus $\left<c_3\right>^2$, so we're really interested in $\left<c_3^2\right>$. Following Ref.\ \cite{Meyer}, recall that a chain of length $3$ requires an integration over two points, with one to the future of the other; thus, $\left<c_3^2\right>$ can be calculated with the integration over the positions of four points, two sets of points with one to the future of the other. There is one caveat: one must proceed cautiously because these points can coincide. Let's label the four points $i$, $j$, $k$, $\ell$, with $j\succ i$ and $\ell\succ k$, to make it easier to keep track. We can have six cases:
\begin{enumerate}
\item All four points are distinct;
\item One coincidence: $i=k$;
\item One coincidence: $i=\ell$;
\item One coincidence: $j = k$;
\item One coincidence: $j = \ell$;
\item Two coincidences: $i = k, \ell = j.$
\end{enumerate}
We can set up integrals for each of these cases; they all proceed trivially with the exception of the fifth case until one notices that its result should be the same as in the second case. We should also note that with our density correction, the first case is \textit{not} equivalent to $\left<c_3\right>^2$. Putting this together yields
\beq
\sigma^2_3
= \left(\frac{(N-2)(N-3)}{N(N-1)}-1\right)\left<c_3\right>^2+2\left<c_4\right>
+ \left<c_3\right>+\frac{2N(N-1)(N-2)}{V^3}\int\text{d}^dx_1
\left(\int\text{d}^dx_2\right)^2,
\eeq
which fits perfectly with simulations. As mentioned above, the others are calculable, but the difficulty increases dramatically with $k$, and there doesn't appear to be a general formula. It seems to be a better option to calculate these from simulations; as long as we use a large number of sprinkled causal sets, this should be fine. Results of such simulations are shown in Fig.\ \ref{sigma}. At its maximum at $k=13$, the square lattice's distribution is nearly $15\,\sigma$ away from the theoretical distribution; thus, any reasonable measure of the closeness of these distributions should conclude that the square lattice is \textit{not} manifoldlike. Alternatively, we could have used the peak and width of the distributions to eliminate the square lattice as a manifoldlike causal set. Recall the chain distributions of manifoldlike causal sets appear to be similar to Gaussians; this means we can characterize them with only two numbers: the peak of the distribution and its width.\footnote{Here we take width to mean full width at half maximum.} We can find the peak in the normal way, by differentiating with respect to $k$ and setting this function equal to zero. Doing this yields the following equation which may be numerically solved for $k_{\text{max}}$.
\beq
   \log\left(\frac{\Gamma(d+1)}{2}\right)-\frac{1}{k_\text{max}-1}
   +\psi^0(N-k_\text{max}+2)-\frac{d\psi^0((k_\text{max}-1)d/2)}{2}
   -\frac{d\psi^0(k_\text{max}d/2)}{2} = 0\;,
\eeq
where $\psi^0(x) =\Gamma(x)^{-1}\,\dd\Gamma(x)/\dd x$. Similarly, we can find the width using this value $k_\text{max}$ and numerically solving this equation:
\bea
   & &\frac{1}{2}\left(\frac{\Gamma(d+2)}{2}\right)^{\!k_\text{max}-2}
   \frac{1}{\Gamma(N-k_\text{max}+2)(k_\text{max}-1)\,
   \Gamma((k_\text{max}-1)d/2)\,\Gamma(k_\text{max}d/2)} \notag\\
   & & = \left(\frac{\Gamma(d+2)}{2}\right)^{\!k-2}
   \frac{1}{\Gamma(N-k+2)(k-1)\,\Gamma((k-1)d/2)\,\Gamma(kd/2)}\;,
\eea
for which there will be two solutions, $k_1$ and $k_2$, with the width being $\Delta:= |k_2-k_1|$. The quantities $(k_\text{max},\Delta)$ should be the average peak and width of chain distributions of causal sets embedded in $d$-dimensional Minkowski space. If we wanted to use these to develop a secondary criterion for manifoldlikeness, we could proceed as we did above and numerically calculate the variations of each of these quantities; however, these will contain the same information as the $\sigma_k^2$'s, so we won't do it here.

For comparison, let's consider the chain length distribution for the 3D lattice whose path length distribution was shown in Fig.\ \ref{fig:3Dlat}. In the case of chains, the separation between this distribution and the corresponding ones for random 3D Minkowski causal sets is greater than in the case of paths, but once again we can't say definitively that this criterion will reject all non-manifoldlike causal sets; however, it will at the very least reject most non-manifoldlike causal sets.

\begin{figure}[H]
\centering
\includegraphics[width = .45\textwidth]{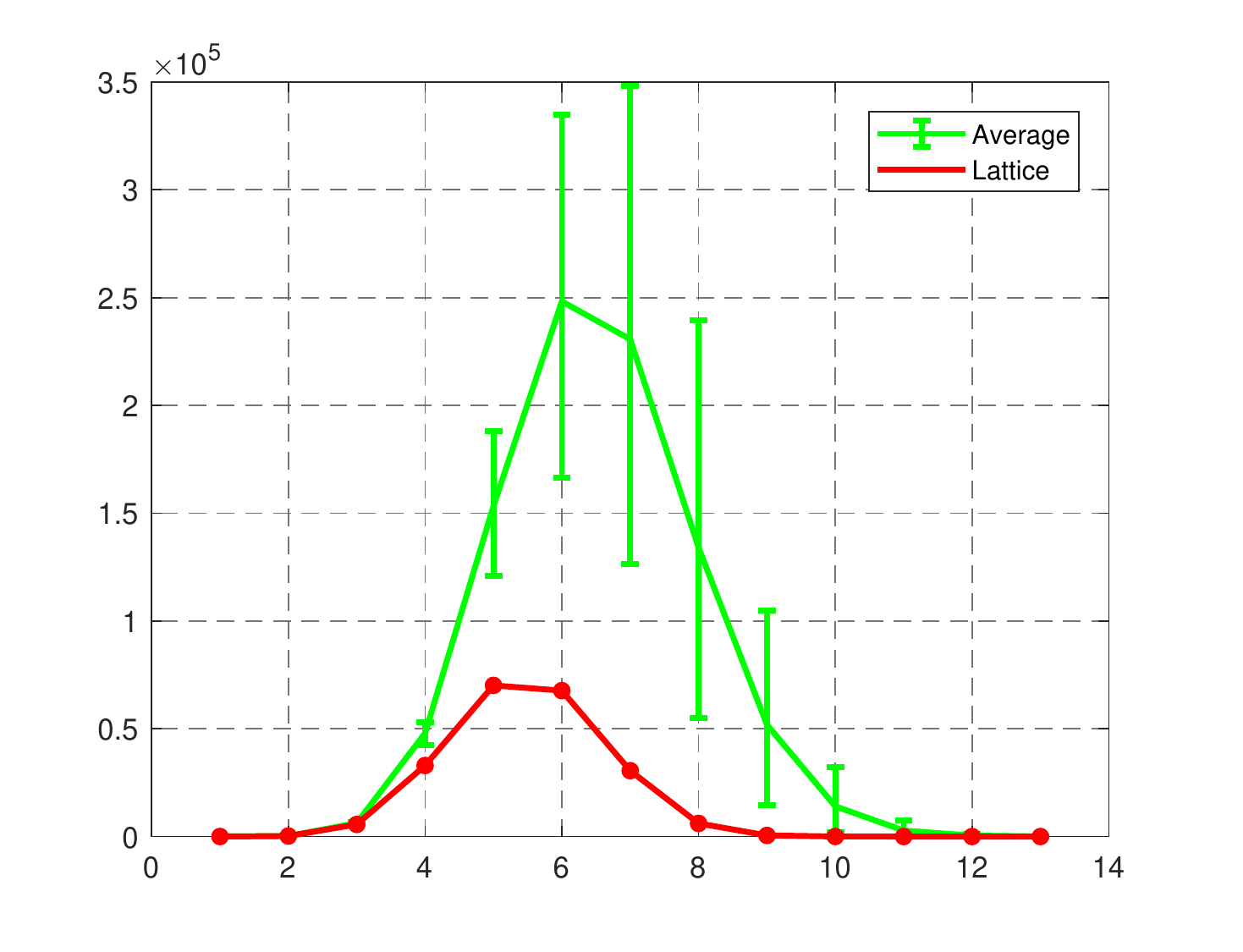}
\caption{\label{fig:3Dlat_paths} Comparison between the chain length distribution for the 3D square lattice and the average of the corresponding distributions for 500 random $235$-point causal sets in 3D Minkowski space; the error bars indicate the range of values corresponding to the 50th percentile above and below the average.}
\end{figure}

\section{Example: Dimension Estimation}

In the previous sections we studied the distributions of chain lengths and path lengths for causal sets uniformly embedded in Minkowski spacetime. We also argued that for a general causal set these distributions are detailed enough to provide a lot of information on its embeddability and, if manifoldlike, on the properties of the Lorentzian manifolds in which they can be embedded. As a first application and example of this statement, we consider here how to use these distributions to determine the dimensionality of a flat spacetime manifold in which a manifoldlike causal set is uniformly embedded.

A number of definitions of dimension for posets are available. The ones that are of interest here are ones that when applied to posets obtained from uniform sprinklings of points in $d$-dimensional Minkowski space give $d$ as the result. The definition of dimension of a poset normally used in combinatorics \cite{KT} does not have this property, except when $d = 2$. The available definitions that do have this property are all probabilistic in nature; can be used to determine the effective dimensionality of different regions within a poset and at different scales, in general with varying results; and come in two kinds. In the first kind \cite{Bthesis,Meyer,BG,Reid}, local neighborhoods are identified with intervals $A(p,q)$ within which combinatorial properties of the poset are analyzed, and their volume (cardinality) is used as a measure of the size of each neighborhood. The second kind of definition, that of spectral dimension \cite{EM,EMS,Carlip}, uses a diffusing particle that executes a random walk starting from a given poset element and following links between elements to identify a local neighborhood around that point, and the scale is determined by the number of steps the particle is allowed to take.

Because these definitions are probabilistic, they are subject to possible systematic and statistical errors, in particular for small posets. In some cases the systematic errors can be modeled and corrected for, but in general neither type of error is worrisome if they are used only for posets uniformly embedded in Minkowski space. Our point of view, however, is that this work is the first step toward addressing the question of recovering useful information from more general posets, ones that may arise from sprinklings in curved manifolds or non-manifoldlike ones. In those settings any noninteger values obtained for the dimensionality may well not be due to the same systematic effects or statistical fluctuations, and it becomes important to identify an approach to dimensionality that in Minkowski space is accurate even when applied to small regions, and that can possibly be modeled to maximize its precision.

In this work we choose to compare the approaches of the first kind above, because with them we do not need to rely on a cutoff as in the spectral dimension approaches, we have started developing an understanding of the behavior of the estimated dimension for curved-geometry and non-manifoldlike posets in one specific approach \cite{paths}, and it appears that the statistical errors are smaller than in approaches of the second kind---though this last point deserves to be studied in more detail.

When trying to recover from a random poset in Minkowski space the dimensionality of the manifold, we need to keep in mind that any meaningful quantity to be extracted from a poset, such as the value of $d$, must be some function of its invariants, i.e., quantities that can be defined and calculated (counted) independently of any labeling of its poset elements; for example, the number of relations, the number of chains of length $k$, the length $k_{\rm max}$ of the longest chain, the number of intervals of cardinality $n$ within it, etc. This leaves many possibilities open, because for a given poset cardinality $N$ most such quantities depend on the dimensionality of the space. Several methods to estimate the Minkowski dimensionality of posets using intervals to identify local neighborhoods have been proposed. In the rest of this section we will compare four of those methods and address the following question: Given a poset that is assumed to be embeddable uniformly in an Alexandrov set of Minkowski space, which of these methods would be the most reliable one for estimating its dimensionality?

Our main criterion for comparing the various approaches will be based on accuracy (rather than precision, for reasons that we will explain below) using numerical simulations and posets obtained from sprinklings in flat space. Specifically, we will generate Minkowski posets by sprinkling $N$ points at random inside an Alexandrov set $A(p,q)$ in $d$-dimensional Minkowski space; $p$ and $q$ themselves will be labeled $p\sb0$ and $p\sb{N+1}$ in the discrete poset; and the latter will therefore be of the form $\mathcal{C} = \{p_i\mid i = 0, ..., N+1\}$ with $\{p\sb1, ..., p\sb{N}\} = A(p\sb0,p\sb{N+1})$.

From a computational point of view one can represent such a poset as one of two types of $(N+2)\times(N+2)$ matrices of 0s and 1s, the matrix of relations {\bf R} or the link matrix {\bf L}, whose elements are defined by
\beq
    R_{ij} = \begin{cases}
    1 & {\rm if}\ p_i \prec p_j \\ 0 & {\rm otherwise} \end{cases}\;,\qquad
    L_{ij} = \begin{cases}
    1 & {\rm if}\ p_i \precc p_j \\ 0 & {\rm otherwise} \end{cases}\;.
\eeq
The two matrices contain the same amount of information and can be determined from each other:
\beq
   R_{ij} = \begin{cases}
   1\hspace{1cm}\text{if $\sum\limits_{n=1}^N (L^n)_{ij}\geq 1$}\\
   0\hspace{1cm}\text{otherwise}\end{cases}
\eeq
and
\beq
   L_{ij} = \begin{cases}
   1\hspace{1cm}\text{if $R_{ij} = 1$ and $(R^2)_{ij}=0$}\\
   0\hspace{1cm}\text{otherwise}\end{cases}.
\eeq 
Notice that the entries in both matrices are always mostly 0s, since $R_{ii} = L_{ii} = 0$ for all $i$, and $(R_{ij} = 1) \Rightarrow (R_{ji} = 0)$ for all $(i,j)$, with $L_{ij}$ satisfying an analogous property; also, for each poset of this type there exists an integer $l\le N+1$ such that ${\bf R}^l = {\bf L}^l \ne {\bf0}$ but ${\bf R}^{l+1} = {\bf L}^{l+1} = {\bf0}$ as matrices. Notice that from the powers of {\bf L} one can read off the number of paths of any given length, and from the powers of {\bf R} one can read off the number of chains of any given length\footnote{Note, this only works if the irreflexive version of the partial order is used.}; more precisely, for any $k \ge 1$ the number of paths/chains of length $k$ between any two $p_i$ and $p_j$ are
\beq
	n^{~}_k(i,j) = (L^k)\sb{ij}\;,\qquad
	c^{~}_k(i,j) = (R^k)\sb{ij}\;, \label{links}
\eeq
respectively.

\begin{figure}[h]
\centering
\includegraphics[width=0.70\textwidth]{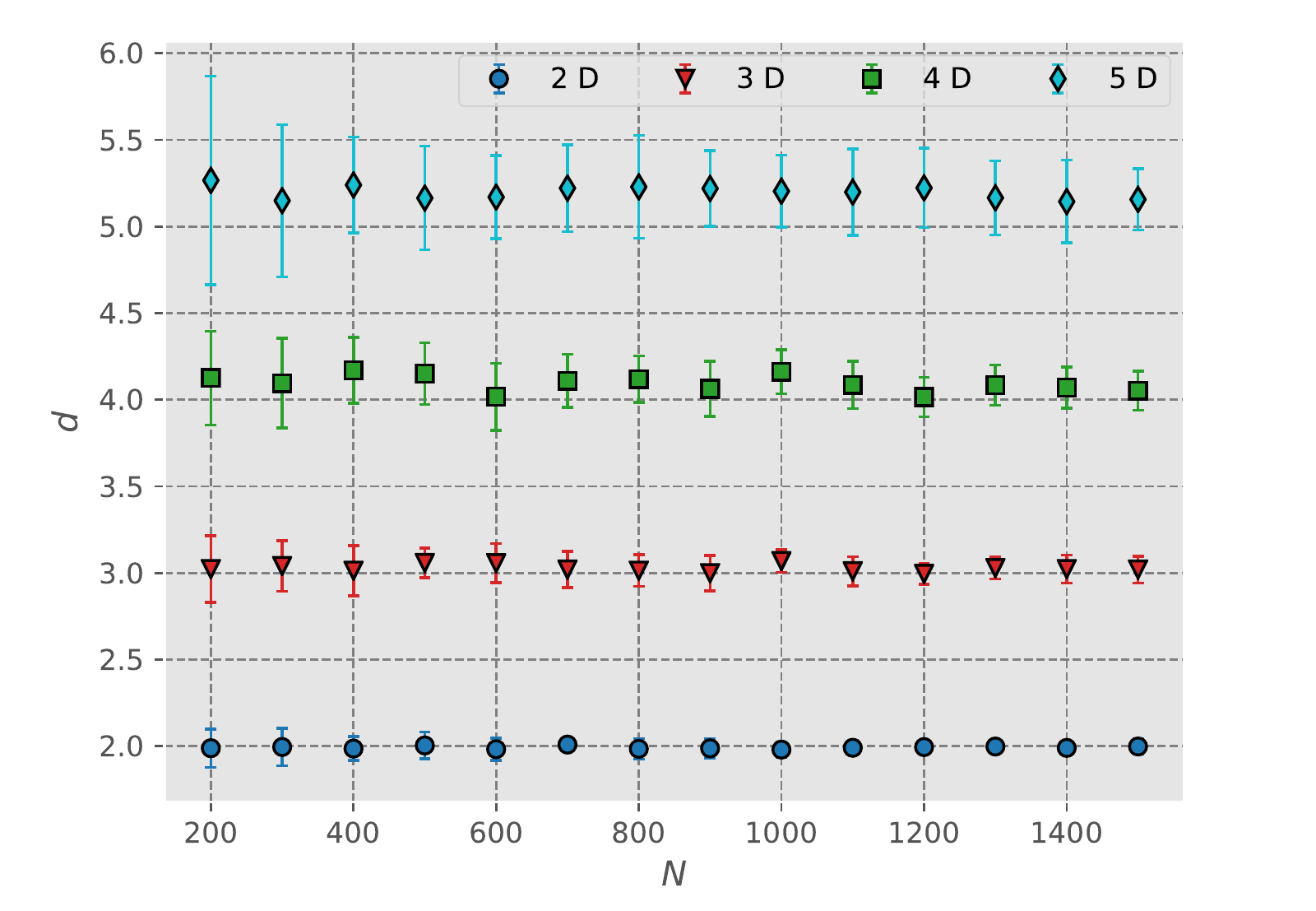}
\caption{\label{dim_mid} The estimated dimension of Minkowski space calculated using the midpoint approach from simulations in $d = 2$, 3, 4 and 5 dimensions as a function of $N$. The error bars indicate the standard deviation of the results obtained from 20 sprinklings generated for each $d$ and $N$.}
\end{figure}

\subsection{Midpoint Approach}
The midpoint dimension of a poset \cite{Bthesis} is obtained as follows. Choose a point $p_m$ in $\mathcal C$ such that the cardinality $n$ of the smaller of the two resultant intervals $I_\downarrow = I(p,p_m)$ and $I_\uparrow = I(p_m,q)$ is as large as possible. (Alternatively, we could choose a point such that the larger of the resulting intervals is as small as possible; simulations we performed show that in that case, or if we used a symmetric combination of the two, the results would be equivalent.) If the overall $N$ is large enough, then this point is most likely in the middle of the big interval, in the sense that the two smaller intervals will have approximately equal volumes and they will be the largest ones for which this is the case. If $\mathcal C$ was uniformly embedded in a flat continuum spacetime, the height of each of $I_\downarrow$ and $A_\uparrow$ would then be approximately half of that of $\mathcal C$ and, because their volumes $V$ scale as the $d$-th power of their height,
\beq
	v = \frac{V}{2^d} \;, \label{v_mid}
\eeq
where $V$ is the volume of $I(p,q)$, and $v$ the volume of the smaller of $I_\downarrow$ and  $I_\uparrow$, in the continuum. But in a poset obtained from a Poisson sprinkling of points in a manifold, the number $N$ of poset elements in each region is proportional to its volume, $N = \rho V$, where $\rho$ is the sprinkling density. As a result, in this approach the Minkowski dimension of the set can estimated as
\beq
	d = \frac{\ln(N/n)}{\ln(2)} \;. \label{d_mid}
\eeq
Fig.\ \ref{dim_mid} shows the estimated dimension, calculated using Eq.\ \ref{d_mid}, for values of $N$ in the range $100\le N\le 1500$ in $d = 2$, 3, 4 and 5 dimensions. For each value of $d$ and $N$, 15 different Minkowski posets were obtained from random point sprinklings in flat spacetime; the error bars show the standard deviations for the 15 corresponding $d$ estimates. What the figure shows is that, in addition to statistical fluctuations that decrease in size for large $N$ and grow for large $d$, the use of Eq.\ \ref{d_mid} introduces a systematic error that also decreases with $N$ and increases with $d$. At least part of this error is due to the fact that because of how it is defined, $v$ is in general smaller than the value given by Eq.\ \ref{v_mid}.

\begin{figure}[h]
\centering
\includegraphics[width=0.70\textwidth]{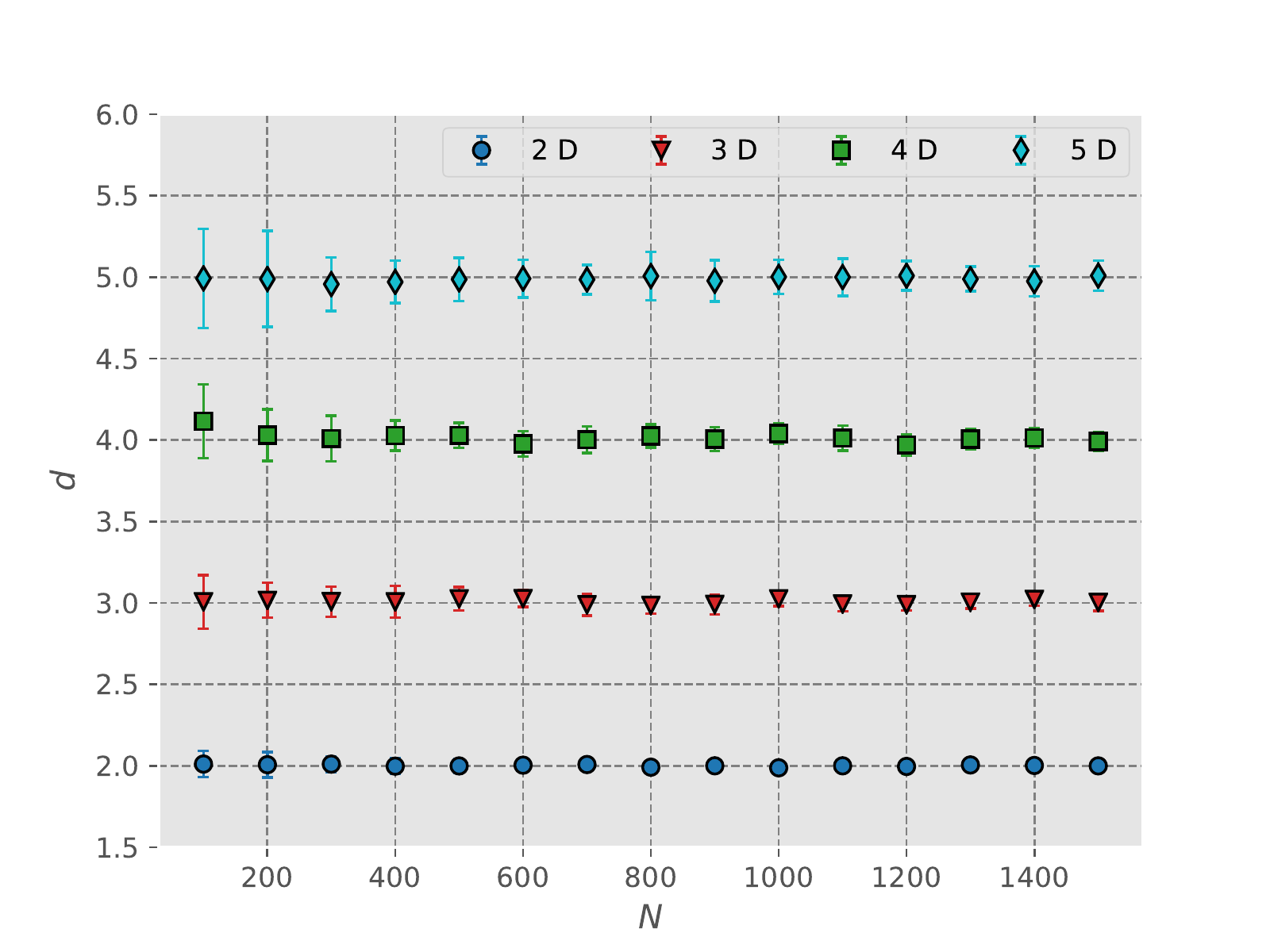}
\caption{\label{dim_MM} The estimated dimension of Minkowski space calculated using the Myrheim-Meyer approach from simulations $d = 2$, 3, 4 and 5 dimensions as a function of $N$. The error bars indicate the standard deviation of the results obtained from 20 sprinklings generated for each $d$ and $N$.}
\end{figure}

\subsection{Modified Myrheim-Meyer Approach}

In Ref.\ \cite{Meyer} Meyer proposed a way to estimate the dimension from the expected number of chains, known as the Myrheim-Meyer dimension; to use it, one calculates the number of 3-chains in a causal set and inserts this into the equation for the expected number of 3-chains:
\beq
	\frac{\left<c_3\right>}{N^2}
	= \frac{1}{4}\,\frac{\Gamma(d+1)\,\Gamma(d/2)}{\Gamma(3d/2)} 
\eeq
which can be solved numerically for $d.$ Our modification to the number of chains improves this equation; however, as only chains of length 3 are used and our correction is smaller for chains of smaller length, the Modified Myrheim-Meyer dimension only slightly deviates from the original particularly for large causal sets. Using our equation,
\beq
\frac{\left<c_3\right>}{N(N-1)} = \frac{1}{4}\frac{\Gamma(d+1)\Gamma(d/2)}{\Gamma(3d/2)}. \label{d_MM}
\eeq
Fig.\ \ref{dim_MM} shows the estimated dimension, calculated using Eq.\ \ref{d_MM}, for values of $N$ in the range $100\le N\le 1500$ in $d = 2$, 3, 4 and 5 dimensions. For each value of $d$ and $N$, 15 different Minkowski posets were obtained from random point sprinklings in flat spacetime; the error bars show the standard deviations for the 15 corresponding $d$ estimates. Because this estimate relies on the exact relation (\ref{d_MM}), rather than on an approximate relationship between discrete and continuum quantities, the figure only shows statistical fluctuations, that again decrease in size for large $N$ and grow with $d$.

\subsection{Brightwell-Gregory Approach}\label{sec_BG}
\begin{figure}[H]
\centering
\begin{minipage}{.45\textwidth}
\centering
\includegraphics[width = \textwidth]{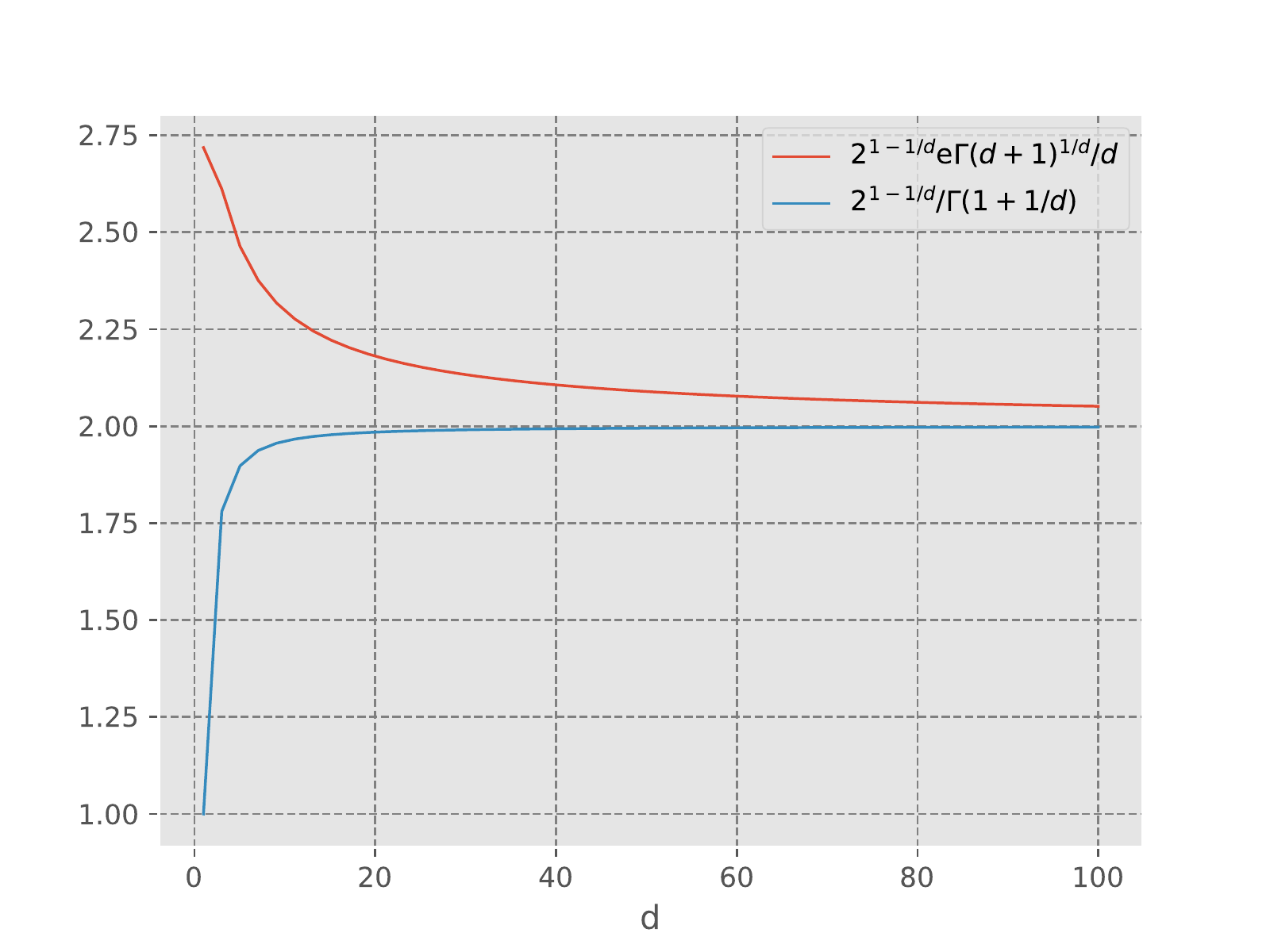}
\end{minipage}
\begin{minipage}{.45\textwidth}
\centering
\includegraphics[width = \textwidth]{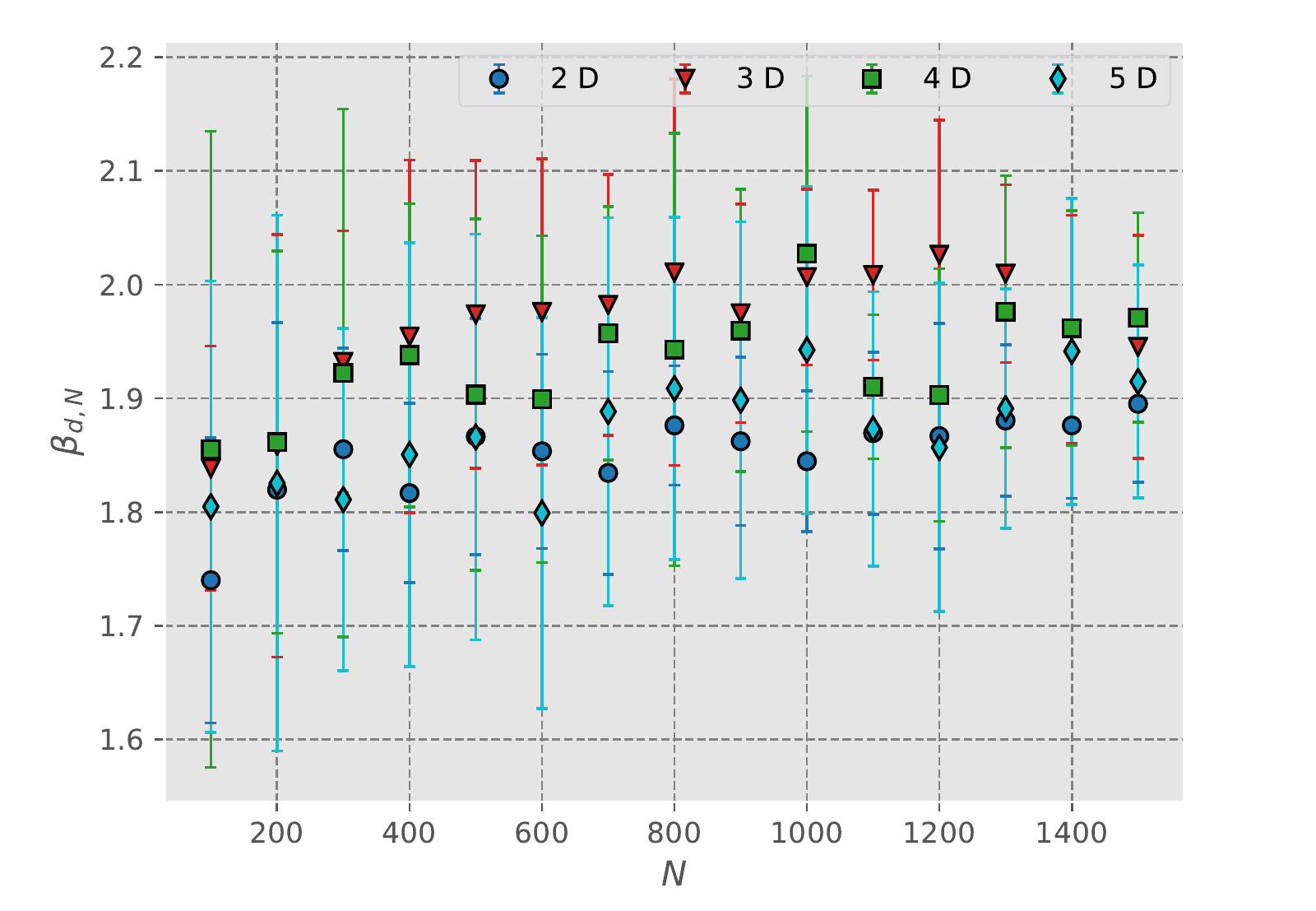}
\end{minipage}
\caption{\label{beta} Left: Upper and lower bounds on the Brightwell-Gregory parameter $\beta_d$ as functions of $d$. Right: The value of $\beta_{d,N}$ for various dimensionalities as a function of $N$. The error bars indicate the standard deviation of the results obtained from 15 sprinklings for each $d$ and $N$.}
\end{figure}

The Brightwell-Gregory method \cite{BG} for calculating the dimension of a Minkowski poset is based on the length of the longest path in the poset. If an $N$-element poset is uniformly sprinkled inside an Alexandrov set of volume $V$ in Minkowski space, the longest path (longest chain) in it can be thought of as a discretized geodesic between the maximal and minimal points. Its length $k_{\rm max}$ should then be approximately proportional to the proper time from $p\sc0$ to $p\sb{N+1}$, which in the continuum is in turn proportional (with a known proportionality coefficient) to $V^{1/d}$. This leads to defining a coefficient $\beta_{d,N}$ by
\beq
	k\sb{\rm max} = \beta_{d,N}\,N^{1/d}\;. \label{kmax}
\eeq
From simulations, it appears that $\beta_{d,N}$ is only weakly dependent on $N$ and in the large-$N$ limit, when endpoint effects become unimportant, it is expected to approach a well-defined value
\beq
    \beta_d:= \lim_{N\to\infty} \beta_{d,N}\;.
\eeq
Brightwell and Gregory found \cite{BG} that in two dimensions $\beta_2=2$, and for arbitrary dimensionality $d\geq 3$ we have
\beq
	1.77\leq \frac{2^{1-1/d}}{\Gamma(1+1/d)} \leq \beta_d\leq
	\frac{2^{1-1/d}\,{\rm e}\,\Gamma(d+1)^{1/d}}{d} \leq 2.62 \;.
\eeq
These upper and lower bounds on $\beta_d$ are plotted as functions of $d$ in Fig.\ \ref{beta}. We can both infer from such plots and find analytically that $\beta_d \to 2$ as $d \to \infty$, and these results are consistent with $\beta_d = 2$ for all $d$, but we are not aware of results to this effect. It is interesting to note that, while these bounds allow a greater range of values for $\beta_d$ above 2 than below 2, it appears from the results of our simulations, shown in Fig.\ \ref{beta}, that for low values of $N$ all $\beta_{d,N}$ are below 2. Those results were obtained generating, for each $d$ between 2 and 5 and values of $N$ between 100 and 1500, 15 random Minkowski posets, finding the length $k_{\rm max}$ of the longest chain in each of them, and averaging the 15 estimated values of $\beta_{d,N}$ obtained from Eq.\ \ref{kmax}.

\begin{figure}[h]
\centering
\includegraphics[width=0.70\textwidth]{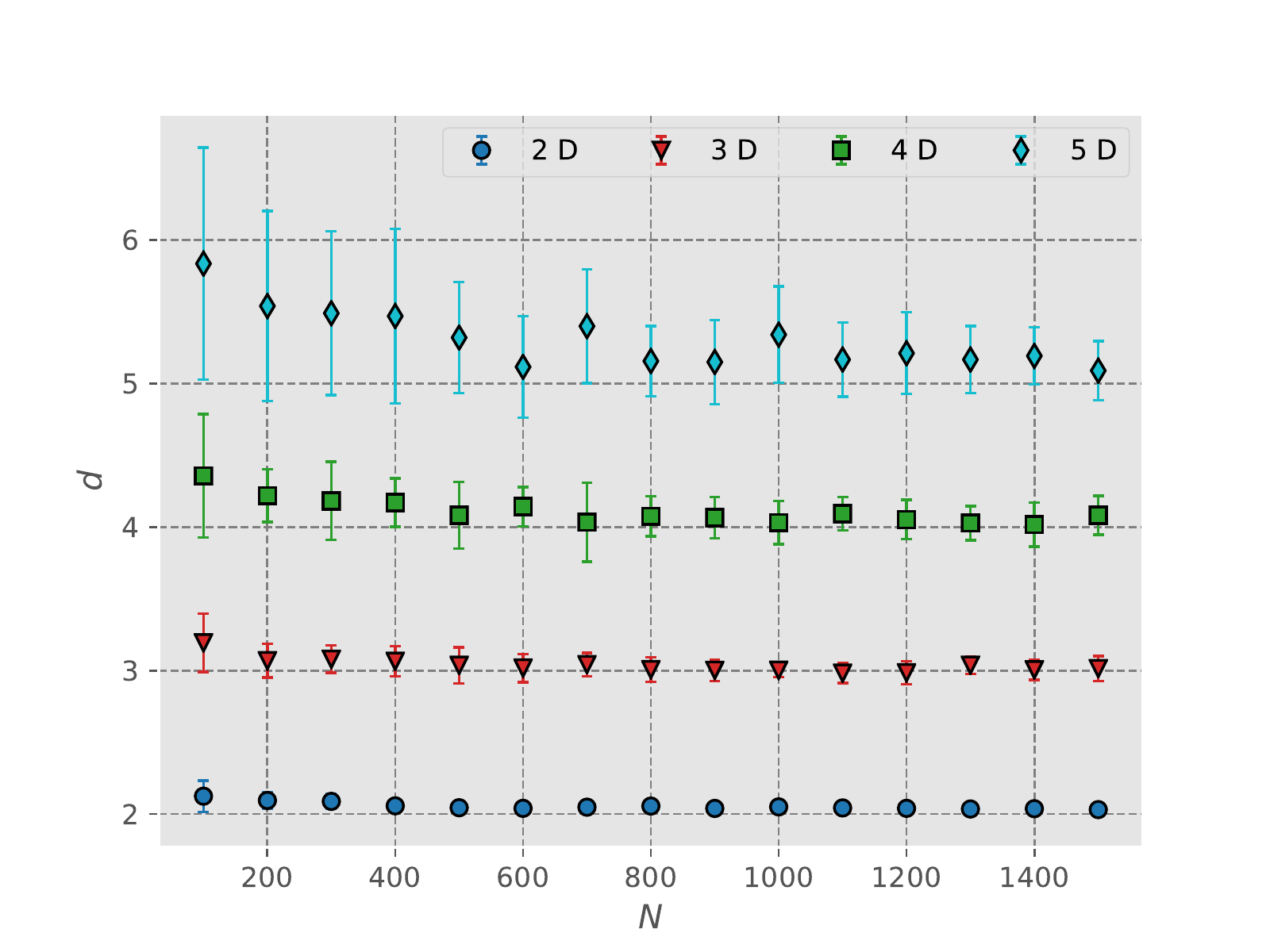}
\caption{\label{dim_BG} The estimated dimension of Minkowski space in the Brightwell-Gregory approach from simulations in $d = 2$, 3, 4 and 5 dimensions as a function of $N$. The error bars indicate the standard deviation of the results obtained from 20 sprinklings for each value of $d$ and $N$.}
\end{figure}

In this approach, the Minkowski dimension of a given poset can be estimated by finding $k_{\rm max}$ and solving for $d$ the defining relation for $\beta_{d,N}$, that can be written as
\beq
	d = \frac{\ln(N)}{\ln(k_{\rm max}/\beta_{d,N})} \;. \label{dim_beta}
\eeq
The most accurate estimate for $d$ would be obtained if we had a theoretical value for $\beta_{d,N}$ available for the given $N$ and for all $d$ in some range, and found a $d$ that solves Eq.\ \ref{dim_beta} exactly, possibly using numerical methods. In the absence of such an exact expression we could produce approximate values of $d$ centered around the correct ones using for $\beta_{d,N}$ the average of the values obtained from some set of simulations, for example the ones shown in Fig.\ \ref{beta}, for each pair of values of $(d,N)$. This would improve the precision of the estimates for $d$ but leave their accuracy essentially unaffected, so we chose not to do it and use $\beta = 2$ in all estimates, which explains the offset of the results shown in Fig.\ \ref{dim_BG}, obtained with the same set of values for $(d,N)$.

\subsection{Average Path Length Approach}

The approach we proposed in Ref.\ \cite{paths} to estimating the Minkowski dimension of a poset is based on the distribution of path lengths in it. Examples of such distributions were shown in Fig.\ \ref{fig:pathscompare} for posets obtained from Minkowski space sprinklings. In these examples the curves are bell-shaped, with an average length $\bar k$ and full width at half maximum $\Delta$ that differ for different values of $d$; if the number of points is kept constant, as the dimension increases both $\bar k$ and $\Delta$ tend to smaller values because of the extra volume introduced by the extra dimensions; the shape of the distribution however remains the same for all dimensions. When estimating $d$, this encourages us to use the average path length $\bar k$, as a quantity that may be subject to smaller statistical fluctuations than $k_{\rm max}$.

The average path length in a poset sprinkled in Minkowski space can be numerically calculated as follows. We first find the total number of $k$-element paths $n\sb{k}$ and the total number of paths $N_{\rm paths}$ between the minimal element $p\sb0$ and the maximal element $p\sb{N+1}$ using the link matrix,
\beq
    n\sb{k} = (L^k)\sb{0,N+1}\;,\qquad N_{\rm paths} = \sum_{k=1}^{N+1} n_k\;.
\eeq
Notice that $N+1$ is the longest possible path length (in fact, $k = N+1$ is only achieved when the poset is totally ordered), but in practice in a simulation the summation for $N_{\rm paths}$ can be stopped after $k = l$ if it is found that $(L^l)\sb{ij} = 0$ for all $(i,j)$. Using these quantities, the average path length can then be found simply as
\beq
    \bar k = \frac1{N_{\rm paths}} \sum_{k=1}^{N+1} k\,n\sb{k}\;.
\eeq

\begin{figure}[h]
\centering
\includegraphics[width=0.70\textwidth]{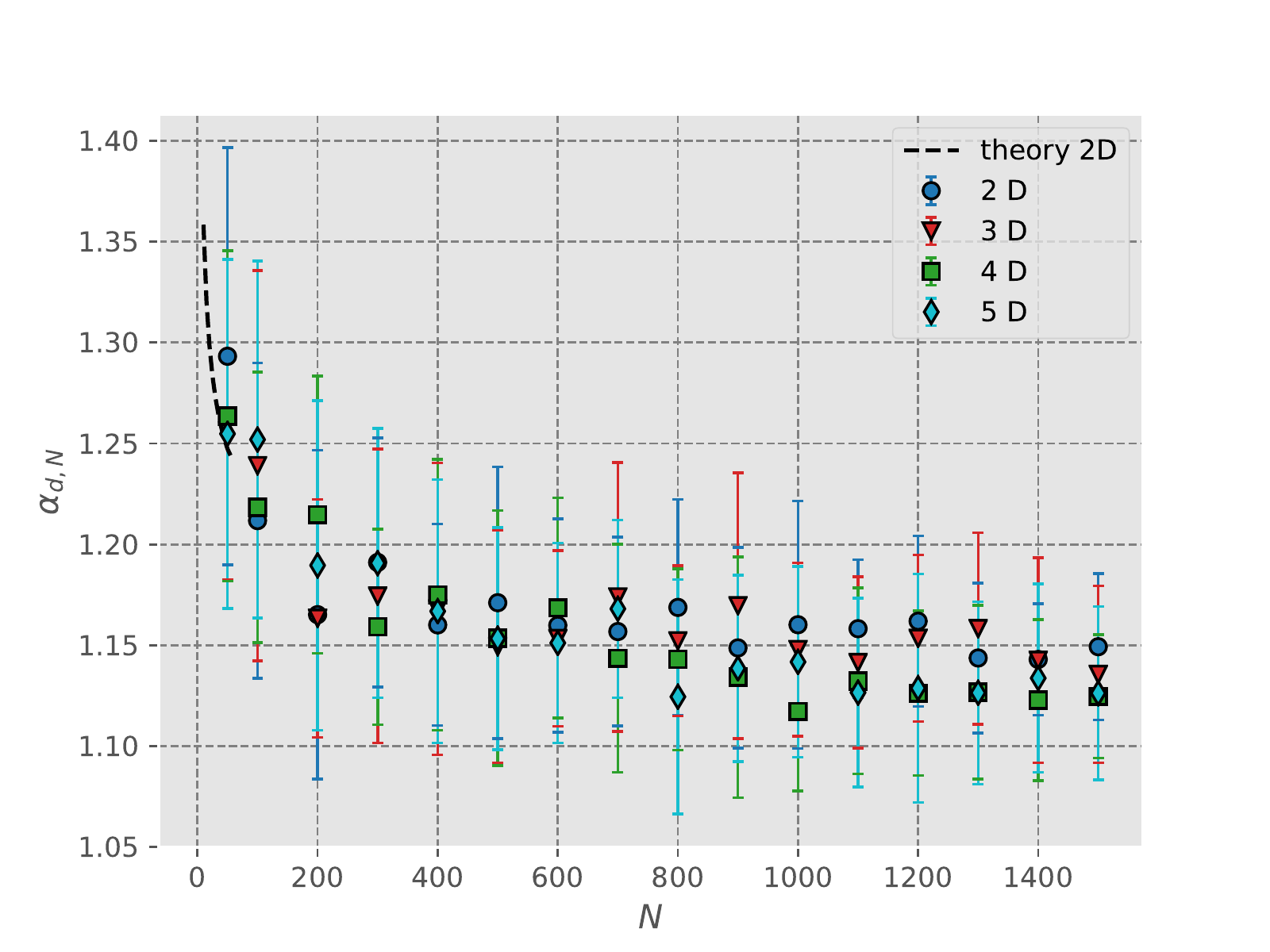}
\caption{\label{alpha} The value of $\alpha_{d,N}$ for various dimensionalities as a function of $N$. The error bars indicate the standard deviation of the results obtained from 20 sprinklings for each $d$ and $N$. The dashed line indicates the analytically calculated values for $d = 2$ and $N \le 55$. For larger values of $N$, we lose the ability to calculate the theoretical values; however, the dashed line shows good agreement with the simulations and extends the results to smaller values of $N.$}
\end{figure}

As for $k_{\rm max}$ in the Brightwell-Gregory approach, in the large-$N$ limit the average path length in any given number of dimensions will be proportional to $N^{1/d}$, so we can write down an analogous relation between $\bar{k}$ and $N$, of the form
\beq
	\bar{k} = \alpha\sb{d,N}\, N^{1/d}\;. \label{kbar}
\eeq
An interesting aspect of our approach is that it is in fact possible in principle to numerically calculate exact values for $\alpha\sb{d,N}$ \cite{paths}. We could then rewrite Eq.\ \ref{kbar} as
\beq
	d = \frac{\ln(N)}{\ln(\bar{k}/\alpha_{d,N})} \;, \label{dim_alpha}
\eeq
and, using the theoretically derived values of $\alpha\sb{d,N}$, find the $d$ that solves Eq.\ \ref{dim_alpha} exactly, at least in principle and again possibly using numerical methods. The problem is that the size of the numbers used to calculate those exact values is a fast growing function of $N$. For $d = 2$ we have only calculated the exact $\alpha_{d,N}$ up to $N = 55$. Fig.\ \ref{alpha} shows those values (dashed line), together with approximate values estimated using 15 Minkowski space sprinklings in $d = 2,3,4,5$ and various values of $N$ up to 1500. The error bars shown indicate the standard deviation of the results obtained from the 15 sprinklings generated for each $d$ and $N$, and are mostly due to statistical fluctuations in the value of the average $\bar k$. (For reference, the computing time for each 1000-point sprinkling with $d = 2$ on a regular laptop computer with a 2.3-Gz Intel Core i5 processor is less than a minute, a number that can be better appreciated if we consider that the total number of paths in the resulting poset is $N_{\rm paths} \approx 3 \times 10^{16}$.)

\begin{figure}[h]
\centering
\includegraphics[width=0.70\textwidth]{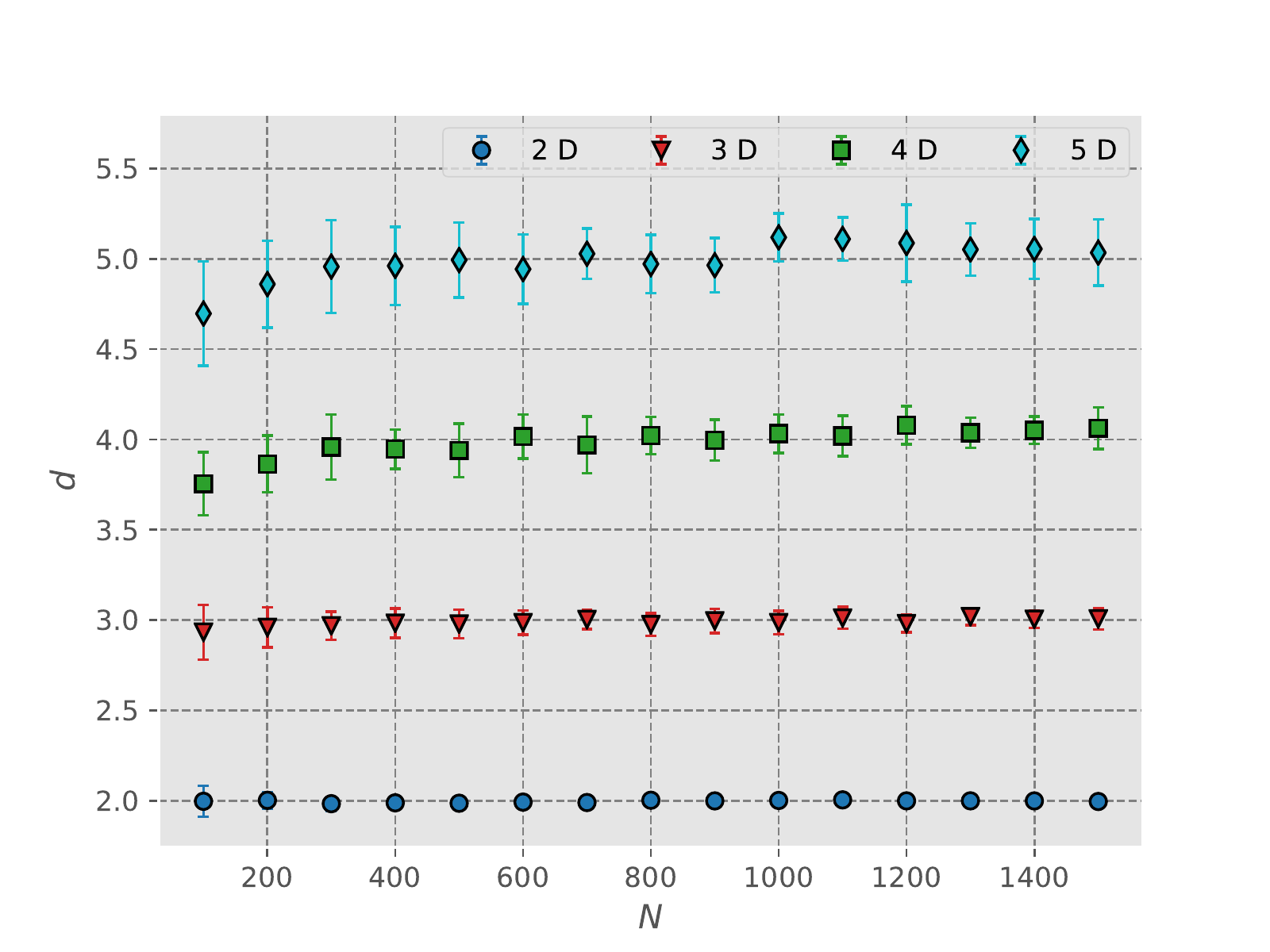}
\caption{\label{dim_ABP} The estimated dimension of Minkowski space calculated using the average path length approach as a function of $N$, from simulations in $d = 2$, 3, 4 and 5 dimensions. The error bars indicate the standard deviation of the results obtained from 20 sprinklings generated for each $d$ and $N$.}
\end{figure}

An interesting observation is that, based on Fig.\ \ref{alpha}, $\alpha\sb{d,N}$ appears to be very weakly dimension dependent and if we define $\alpha\sb{N} \equiv \alpha\sb{d,N}$, then $\alpha\sb{N}$ appears to tend to the asymptotic value of 1.15 for large values of $N$. The dimension estimates for the posets generated with the above values for $d$ and $N$ are shown in Fig.\ \ref{dim_ABP}. For reasons similar to the ones explained in Section \ref{sec_BG} for the Brightwell-Gregory approach, the value of $\alpha$ we used in Eq.\ \ref{dim_alpha} to estimate $d$ was 1.15 for all $(d,N)$.

\subsection{Interval Size Approach}

\begin{figure}[h]
\centering
\includegraphics[width=0.70\textwidth]{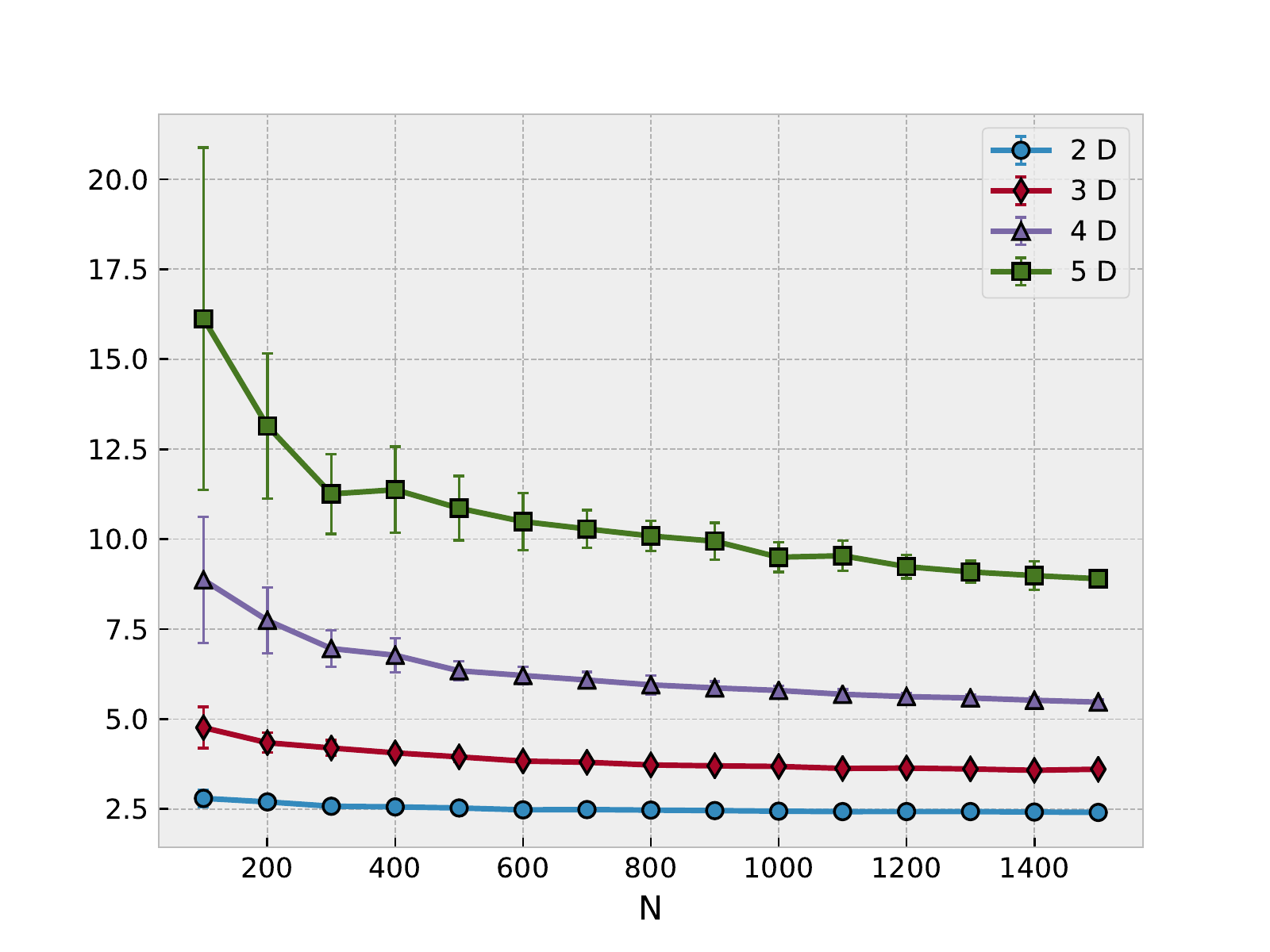}
\caption{\label{intervals} The estimated dimension of Minkowski space calculated using the interval size approach as a function of $N$, from simulations in $d = 2$, 3, 4 and 5 dimensions. The error bars indicate the standard deviation of the results obtained from 20 sprinklings generated for each $d$ and $N$.}
\end{figure}

Finally, it is worth mentioning another article \cite{GlaserSurya} in which the authors calculate the characteristic shape of the distribution of interval sizes, as a tool for defining local and non-local neighborhoods. The distribution has the following form
\bea
&& \left<a_k\right> = \frac{(\rho V_0)^{k+2}}{(k+2)!}\,\Gamma(d)^2\,\frac{\Gamma((k+1)d/2+1)\Gamma(kd/2+1)}{\Gamma(\Gamma((k+1)d/2+d)\Gamma(kd/2+d)}\times \\
&& \kern10pt \times\ _dF_d\left(\{k+1,\tfrac{2}{d}+k, \cdots,\tfrac{2(d-1)}{d}+k\}; \{3+k, \tfrac{2}{d}+k+2, \cdots, \tfrac{2(d-1)}{d}+k+2\}; -\rho V_0\right). \notag
\eea
In addition to the main purpose for writing own this result, the authors obtain an asymptotic formula for the ratio of the number of intervals of size $k$ to that of intervals of size 0, which in the asymtotic limit $\rho\to\infty$ (with $V_0$ fixed) only depends on the dimension,
\beq
\lim_{\rho\rightarrow\infty} \frac{\left<a_k\right>}{\left<a_0\right>} = \frac{\Gamma(2/d+k)}{\Gamma(2/d)\Gamma(k+1)}.
\eeq
Using this formula, however, does not provide an accurate estimate for the dimension, precisely because it is an asymptotic formula which approaches the correct result only in the limit $\rho\rightarrow\infty$, as can be seen from Fig.\ \ref{intervals}.

\section{Conclusions}

The purpose of this paper was to calculate the expected distributions of chain and path lengths for causal sets embeddable in $d$-dimensional Minkowski space, and draw attention to a correction that needs to be introduced to previous calculations of these quantities. We showed that the effects of this correction on the length probability distributions are not negligible, but they can be treated relatively easily. We then showed that these calculations can be used in a comparison between a particular causal set's chain/path length distribution with the expected distribution for a Minkowski-like causal set to determine if the causal set is embeddable in Minkowski space (or, if not, how close it is to being embeddable), and if it is to determine its dimension. Compared to paths, chain lengths have the benefit that their distribution as a whole is known explicitly, and this analytical result makes it simpler to use in different situations.

We also discussed different types of non-manifoldlike causal sets that are dominant in numbers, the Kleitman-Rothschild type causal sets, and the frequently used ones based on regular lattices, and showed that our criterion is precise enough to eliminate them as manifoldlike. While it appears that most non-manifoldlike causal sets are disqualified by this criterion, nonetheless, one can use the whole distribution or the reduced parameter space (peak position, full width at half maximum, and the total number of paths) to define a manifoldlikeness measure for general causal sets. This is important in particular if we are to consider causal sets with some roughness in the microscopic scales.

The above results can be applied to manifoldlike causal sets to obtain their dimension. Even in these general terms, this question is of interest, but it had already been addressed by various satisfactory approaches to dimension estimation, including ones based on exact theoretical calculations. Of the approaches we looked at in our comparison of the results obtained for numerically simulated, uniformly distributed posets in Minkowski space, the best ones in terms of size of statistical fluctuations are the modified Myrheim-Meyer method, based on the number of relations $\left<c_3\right>$, and the method based on the average path length $\bar{k}$. In particular, we were interested in the size of the relative fluctuations in the dimension estimates for small posets. One reason for this is that we view the study of Minkowski posets as a first step in the study of more general manifoldlike posets, and we expect small enough subsets of posets embeddable in curved Lorentzian geometries to be close enough to Minkowski posets that they can be used to obtain dimensional information.

Since the dimensionality of a manifold is an integer number and all or most of the methods described here (with the possible exception of the Brightwell-Gregory approach) show relatively small fluctuations, if all posets we consider were known to be manifoldlike there would be various approaches we could use. However, the vast majority of posets are not manifoldlike (Ref.\ \cite{paths} discusses this point in a little more detail), so we would in fact like to be able to extend our method to some of the non-manifoldlike ones and obtain from it information on the type of obstruction to their embeddability, at least when they are close to being manifoldlike in an appropriate sense. In the latter case, an estimate of $d$ might give a value that is close to, but does not coincide with one corresponding to an integer dimension. This is the reason why both small statistical fluctuations and precise theoretical modeling are important.

\end{document}